\global\def\draftcontrol{0}

   \def\versionno{ holomorphic anomaly }

\catcode`\@=11

\expandafter\ifx\csname draftcontrol\endcsname\relax\global\def\draftcontrol{0} 
\fi 

{\count255=\time\divide\count255 by 60 
\xdef\hourmin{\number\count255} 
\multiply\count255 by-60\advance\count255 by\time 
\xdef\hourmin{\hourmin:\ifnum\count255<10 0\fi\the\count255}} 
\def\draftdate{\number\month/\number\day/\number\year\ \ \ \hourmin } 

\newcommand\makepapertitle{\par

  \begingroup 
    \renewcommand\thefootnote{\@fnsymbol\c@footnote}%
    \def\@makefnmark{\rlap{\@textsuperscript{\normalfont\@thefnmark}}}%
    \long\def\@makefntext##1{\parindent 1em\noindent 
            \hb@xt@1.8em{%
                \hss\@textsuperscript{\normalfont\@thefnmark}}##1}%
     \newpage 
     \global\@topnum\z@   
     \@makepapertitle 
     \thispagestyle{empty}\@thanks 
  \endgroup 
  \setcounter{footnote}{0}%
  \global\let\thanks\relax 
  \global\let\makepapertitle\relax 
  \global\let\@makepapertitle\relax 
  \global\let\@thanks\@empty 
  \global\let\@author\@empty 
  \global\let\@date\@empty 
  \global\let\@title\@empty 
  \global\let\title\relax 
  \global\let\author\relax 
  \global\let\date\relax 
  \global\let\and\relax 
  \def\version{\let\version\@version\@gobble} 
} 
\def\@makepapertitle{%
  \newpage 
   \ifnum\draftcontrol=1 {} 
   \version\versionno 
   \vskip 5em%
   \else 
   \hfill\hbox to 3cm {\parbox{4cm}{\@pubnum}\hss}%
   \vskip 5em%
   \fi 
   \begin{center}%
   \let \footnote \thanks 
      {\hskip -0\textwidth \hbox to 1\textwidth%
        {\centerline{\Large\bf{\noindent\@title}}}}%
     \vskip 1.5em%
     {\normalsize
       \lineskip .5em%
       \begin{tabular}[t]{c}%
         \@author 
       \end{tabular}\par}%
     \vskip 1.5em%
     {\@bstract}%
     \end{center}%
     \vfill
     \@date%
     \vskip 1.5em%
   \par 
} 

\gdef\@pubnum{} 
\def\pubnum#1{%
  \gdef\@pubnum{#1}} 

\gdef\@bstract{} 
\def\Abstract#1{%
  \gdef\@bstract{%
   \parbox{\textwidth-0pc}{%
   \centerline{\bf Abstract}\penalty1000 
   \noindent
   \renewcommand\baselinestretch{1.0} 
   {#1}}} 
} 

\gdef\@email{}
\def\email#1{%
   \gdef\@email{%
   Email: {\tt #1}}
}

\def\ps@paper{\let\@mkboth\@gobbletwo%
     \ifnum\draftcontrol=1 
        \def\@oddfoot{\hbox to \textwidth{\tiny \versionno \hfil\tiny\draftdate}%
        \hskip -\textwidth \hbox to \textwidth{\hfil\rm\thepage\hfil}}%
     \else\def\@oddfoot{\hbox to \textwidth{\hfil\rm\thepage\hfil}} 
     \fi 
     \let\@evenfoot\@oddfoot 
} 

\def\body{\clearpage 
          \pagestyle{paper} 
        } 
\newenvironment{acknowledgments}{%
\vskip 3.25ex 
\addcontentsline{toc}{section}{Acknowledgments}
\noindent {\bf Acknowledgments} 
} 

\def\@version#1{\ifnum\draftcontrol=1 
\typeout{}\typeout{#1}\typeout{} 
\vskip3mm\centerline{\hbox{\fbox{\normalsize{\tt DRAFT -- #1 -- } 
                   {\draftdate}}}}\vskip3mm 
\fi} 
\let\version\@version 
\long\def\eqlabel#1{\ifnum\draftcontrol=1 
                    \tag@false  
                    \tag*{(\theequation) \hbox to -0.2cm{\hspace{0cm}\small{#1}\hss}} 
                    \refstepcounter{equation}  
                    \edef\@currentlabel{\theequation} 
                    \ltx@label{#1}          
                    \else 
                    \label{#1} 
                    \fi 
                    } 
\let\st@bibitem\@bibitem 
\let\st@lbibitem\@lbibitem 
\ifnum\draftcontrol=1 
  \def\@bibitem#1{%
    \st@bibitem{#1}\a@@label{#1}\ignorespaces} 
  \def\@lbibitem[#1]#2{%
    \st@lbibitem[#1]{#2}\a@@label{#2}\ignorespaces} 
  \def\a@@label#1{%
    \gdef\a@lab{\smash{\normalfont\small#1}} 
    \ifvmode 
      \if@inlabel 
        \global\setbox\@labels\hbox{%
          \llap{\a@lab\let\a@lab\relax 
                \kern\@totalleftmargin\kern\marginparsep}%
          \box\@labels}%
      \fi 
    \fi} 
\fi 

\documentclass[12pt,letterpaper]{article} 

\usepackage{amsmath,bm,amsfonts,amssymb,array,calc,amsthm,rotating}
\usepackage{epsfig,psfrag} 
\usepackage{graphicx}
\usepackage{color}
\usepackage[colorlinks=false]{hyperref}

\renewcommand\baselinestretch{1.25} 
\renewcommand\section{\@startsection {section}{1}{\z@}%
                                   {-3.5ex \@plus -1ex \@minus -.2ex}%
                                   {2.3ex \@plus.2ex}%
                                   {\normalfont\large\bfseries}} 
\renewcommand\subsection{\@startsection{subsection}{2}{\z@}%
                                   {-3.25ex\@plus -1ex \@minus -.2ex}%
                                   {1.5ex \@plus .2ex}%
                                   {\normalfont\normalsize\bfseries}} 
\renewcommand\subsubsection{\@startsection{subsubsection}{3}{\z@}%
                                   {-3.25ex\@plus -1ex \@minus -.2ex}%
                                   {1.5ex \@plus .2ex}%
                                   {\normalfont\normalsize\it}} 
\renewcommand\paragraph{\@startsection{paragraph}{4}{\z@}%
                                   {-3.25ex\@plus -1ex \@minus -.2ex}%
                                   {1.5ex \@plus .2ex}%
                                   {\normalfont\normalsize\bf}} 
\renewcommand\subparagraph{\@startsection{subparagraph}{5}{\z@}%
                                   {-1.25ex\@plus -1ex \@minus -.2ex}%
                                   {0ex \@plus .2ex}%
                                   {\normalfont\normalsize\it}}

\numberwithin{equation}{section}

\long\def\@makecaption#1#2{%
  \vskip\abovecaptionskip
  \sbox\@tempboxa{{\bf #1:} #2}%
  \ifdim \wd\@tempboxa >\hsize
    {\small\bf #1:} {\small #2}\par
  \else
    \global \@minipagefalse
    \hb@xt@\hsize{\hfil\box\@tempboxa\hfil}%
  \fi
  \vskip\belowcaptionskip}

\renewcommand*\l@section[2]{%
  \ifnum \c@tocdepth >\z@
    \addpenalty\@secpenalty
    \addvspace{.5em \@plus\p@}%
    \setlength\@tempdima{1.5em}%
    \begingroup
      \parindent \z@ \rightskip \@pnumwidth
      \parfillskip -\@pnumwidth
      \leavevmode \bfseries
      \advance\leftskip\@tempdima
      \hskip -\leftskip
      #1\nobreak\hfil \nobreak\hb@xt@\@pnumwidth{\hss #2}\par
    \endgroup
  \fi}
\renewcommand*\l@subsection{\addvspace{.0em \@plus\p@}\@dottedtocline{2}{1.5em}{2.3em}}
\renewcommand*\l@subsubsection{\addvspace{-.2em \@plus\p@}\@dottedtocline{3}{3.8em}{3.2em}}

\def\hepth#1{\href{http://xxx.arxiv.org/abs/hep-th/#1}{{arXiv:hep-th/#1}}}

\def\math#1{\href{http://xxx.arxiv.org/abs/math/#1}{{arXiv:math/#1}}}

\def\alggeom#1{\href{http://xxx.arxiv.org/abs/alg-geom/#1}{{arXiv:alg-geom/#1}}}
\def\arxiv#1#2{\href{http://xxx.arxiv.org/abs/#1}{{arXiv:#1 [#2]}}}

\definecolor{refcol}{rgb}{0.2,0.2,0.8}
\definecolor{eqcol}{rgb}{.6,0,0}
\definecolor{purple}{cmyk}{0,1,0,0}

\gdef\@citecolor{refcol}
\gdef\@linkcolor{eqcol}
\def\colorlinkspurple{\gdef\@urlcolor{purple}}
\def\colorlinksblue{\gdef\@urlcolor{blue}}
\def\colorlinksred{\gdef\@urlcolor{red}}

\def\ie{{\it i.e.}} 
\def\eg{{\it e.g.}} 
\def\etc{{\it etc.}}
\def\cf{{\it cf.}}

\def\revise#1       {\raisebox{-0em}{\rule{3pt}{1em}}%
                     \marginpar{\raisebox{.5em}{\vrule width3pt\ 
                     \vrule width0pt height 0pt depth0.5em 
                     \hbox to 0cm{\hspace{0cm}{%
                     \parbox[t]{4em}{\raggedright\footnotesize{#1}}}\hss}}}}

\def\cale         {{\cal E}} 
\def\calf         {{\cal F}}

\def\calh         {{\cal H}}

\def\call         {{\cal L}} 
\def\calm         {{\cal M}} 
\def\caln         {{\cal N}} 
\def\calo         {{\cal O}}

\def\calt         {{\cal T}} 
 
\def\calv         {{\cal V}} 
\def\calw         {{\cal W}} 
\def\caly         {{\cal Y}}

\def\complex      {{\mathbb C}} 
 
\def\projective   {{\mathbb P}} 
 
\def\reals        {{\mathbb R}} 
\def\zet          {{\mathbb Z}} 

\def\del          {\partial} 
\def\delbar       {\bar\partial} 
\def\ee           {{\it e}} 
\def\ii           {{\it i}} 
 
\def\tr           {{\rm Tr}}

\newcommand\topa[2]{\genfrac{}{}{0pt}{2}{\scriptstyle #1}{\scriptstyle #2}}

\def\sqr#1#2{{\vcenter{\vbox{\hrule height.#2pt   
 \hbox{\vrule width.#2pt height#1pt \kern#1pt 
 \vrule width.#2pt}\hrule height.#2pt}}}}

\def\ab{{\bar a}}
\def\bb{{\bar b}}

\def\ib{{\bar i}}
\def\jb{{\bar j}}
\def\kb{{\bar k}}
\def\lb{{\bar l}}
\def\mb{{\bar m}}
\def\nb{{\bar n}}

\def\qb{{\bar q}}
\def\zb{{\bar z}}

\def\0b{{\bar 0}}

\def\Cb{{C}}
\def\Db{{D}}

\def\Gb{{\overline G}}
\def\Jb{{\overline J}}
\def\phib{{\bar \phi}}
\def\mub{{\bar \mu}}

\def\F#1#2{{\calf}^{(#1,#2)}}
\def\calm#1#2{{\cal M}^{(#1,#2)}}
\def\f#1#2{f^{(#1,#2)}}
\def\A#1#2#3{A^{(#1,#2)}_#3}
\def\n#1#2#3{n^{(#1,#2)}_#3}

\def\Fc#1{{\calf}^{(#1)}}
\def\calmc#1{{\cal M}^{(#1)}}

\def\Omegab{{\overline \Omega}}

\def\pio{\pi^{\rm orb}}
\def\pic{\pi^{\rm c}}
\def\tauo{\tau^{\rm orb}}
\def\tauc{\tau^{\rm c}}
\def\caltc{\calt^{\rm c}}

\def\callc{\call^{\rm c}}

\def\om{\omega}
\def\xb{{\bar x}}

\def\ch{{\rm ch}}
\def\td{{\rm Td}}

\def\Ipp{\mathord{\mathchar "0271 \kern-4.5pt \mathchar"0271}}

\catcode`\@=12 

\begin{document} 



\title{
\parbox{\textwidth}{\begin{center} Extended Holomorphic Anomaly \\[-.1cm] 
and \\[-.1cm]
Loop Amplitudes in Open Topological String
\end{center} }}

\pubnum{%
arXiv:0705.4098}
\date{May 2007}

\author{
Johannes Walcher \\[0.2cm]
\it School of Natural Sciences, Institute for Advanced Study\\
\it Princeton, New Jersey, USA
}

\Abstract{
Open topological string amplitudes on compact Calabi-Yau threefolds are 
shown to satisfy an extension of the holomorphic anomaly equation of 
Bershadsky, Cecotti, Ooguri and Vafa. The total topological charge of 
the D-brane configuration must vanish in order to satisfy tadpole 
cancellation. The boundary state of such D-branes is holomorphically 
captured by a Hodge theoretic normal function. Its Griffiths' 
infinitesimal invariant is the analogue of the closed string 
Yukawa coupling and plays the role of the terminator in a Feynman 
diagram expansion for the topological string with D-branes. The 
holomorphic anomaly equation is solved and the holomorphic ambiguity 
is fixed for some representative worldsheets of low genus and with 
few boundaries on the real quintic.
}

\makepapertitle

\body

\version\versionno

\vskip 1em

\tableofcontents
\newpage

\section{Introduction and Summary}

The topological phase of string theory is one of the cornerstones
of modern mathematical physics. The theory is highly solvable, perhaps
even completely integrable, while at the same time capturing a wealth
of highly non-trivial mathematical and physical information about 
the vacuum geometry of superstring theory. The topological string 
shares many important dynamical features with the ordinary string, 
including D-branes and gauge/gravity duality. Moreover, topological 
strings on Calabi-Yau threefolds in fact directly compute certain 
higher-derivative F-terms in the effective action of the critical 
superstring compactified on the same manifold. This allows the topological 
string to be an important ingredient in the understanding of string 
duality, as well as in black hole entropy computations.

One of the reasons that the topological string is so efficiently solvable
is the existence of various differential equations satisfied by topological 
correlation functions. These differential equations originate for instance 
from topological Ward identities expressing independence of the worldsheet 
theory from the worldsheet metric. The special structures of the topological 
theory allow to integrate these equations up to some integration constants 
which are specified by classical data of the underlying model.

There are many reasons for which the most interesting construction is to 
topologically twist a family of 
unitary $\caln=2$ superconformal field theories of central charge
$\hat c=3$. Such theories have the richest set of non-trivial topological
amplitudes, and are most directly linked to physics in four dimensions.
In this critical instance, the relevant differential equation is the 
holomorphic anomaly equation of Bershadsky, Cecotti, Ooguri and
Vafa \cite{bcov1,bcov2}, which controls the amplitudes as functions over
coupling space. As will be reviewed more extensively below, the 
holomorphic anomaly is deeply rooted in the unitarity or CPT invariance 
of the underlying $\caln=2$ CFT, which leads to an identification of 
BRST and anti-ghost cohomology. The latter is therefore non-trivial, in 
distinction to say the physical bosonic string. These states, 
although BRST trivial, do not decouple from the topological amplitudes,
but they fail to do so in a very controlled fashion, precisely captured
by the holomorphic anomaly equation. The integration of this equation 
leads to polynomial expressions for all
topological amplitudes in terms of tree-level data plus (at each order
of perturbation theory) a finite number of integration constants,
the so-called holomorphic ambiguity. The tree-level data for the 
critical topological string is a special K\"ahler manifold such as for
instance the moduli space of a Calabi-Yau manifold, first computed in
the work of Candelas et al.\ \cite{cdgp}. Fixing the holomorphic 
ambiguity usually requires physical insights into the space-time physics 
associated with the corresponding compactification of the type II string.

Hitherto, most discussions of the holomorphic anomaly have focused on the
closed string. On the one hand, it can be argued that the closed 
topological string is more intrinsic to the Calabi-Yau geometry as 
it does not depend on the choice of a D-brane configuration on top of 
the manifold. On the other hand, this point of view completely ignores 
the central role played by D-branes in mirror symmetry and of course in 
topological gauge/gravity duality, and hence for closed strings themselves.
Any exploration of these topics does require knowledge of topological
amplitudes in the presence of D-branes, \ie, on worldsheets with
boundaries. So far, these amplitudes have been obtained only 
indirectly, such as from matrix models or via Chern-Simons theory, 
or in rather brute force toric computations in the A-model. It is
clearly desirable to develop a more systematic approach to open
topological string amplitudes, at the loop as well as at tree level.

The computation of Candelas et al.\ \cite{cdgp} was recently extended 
to the open string in ref.\ \cite{open}. In that paper, a differential
equation was proposed whose solution gives the tension of the 
domainwall between two vacua on a certain brane wrapped on the quintic,
as a function of the single closed string modulus. From the topological
string perspective, one is computing the disk amplitude. When expanded 
in A-model variables, this solution contains the number (Gromov-Witten
invariant) of holomorphic disks ending in a non-trivial one-cycle on 
the brane. These predictions were partially checked in \cite{open}, and 
fully verified in \cite{psw}. The B-model origin of the differential 
equation of \cite{open} is being explained in \cite{mowa}, in line with
what had been anticipated in previous work \cite{lmw}.

The consistency of the emerging general picture of the open topological
string at tree level has given hope that the computation could be 
extended to higher worldsheet topologies, analogously to what was 
done in \cite{bcov1,bcov2} for the closed string, by using the 
holomorphic anomaly. We will show in this paper that this is indeed 
possible. The holomorphic anomaly for open string has been previously
studied in \cite{bcov2,agnt2,marcos1,eynard}, and was interpreted 
also in \cite{integrable,anv}. In particular, in \cite{eynard}, open 
topological string amplitudes on certain local Calabi-Yau geometries 
are computed to all orders using matrix models, and found to satisfy a 
holomorphic anomaly equation. The paper \cite{marcos1} discusses open 
topological string amplitudes in local toric geometries more generally, 
and finds an explicit expression for the holomorphic anomaly of the 
annulus amplitude. It will be very interesting to elucidate the relation 
to our present work, which is focused on compact Calabi-Yau manifolds.

We have divided the bulk of the paper into two parts. In the first part
(section 2), we will review from \cite{bcov2} the vacuum geometry of 
twisted $\caln=2$ models as well as the derivation of the holomorphic 
anomaly for closed string amplitudes. In parallel, we will describe 
the extension to open strings. In the second part of the paper (section 3), 
we will apply the general formalism to compute topological amplitudes 
on various worldsheets with boundary on the real quintic, which is the 
D-brane geometry that was solved in \cite{open}. But now, let us 
summarize the main ideas and results of this paper.

\subsection{General theory}

The quantities of interest in this paper are the perturbative topological 
string amplitudes $\F gh$ for open plus closed strings. The $\F gh$ are 
defined by integrating over the moduli space, $\calm gh$, of (oriented) 
Riemann surfaces of genus $g$ and with 
$h$ boundary components, the appropriate correlation function of 
the topologically twisted 2d worldsheet theory. The $\F gh$ are 
functions (or rather, sections of an appropriate bundle) over coupling 
space, $M$, which is a complex manifold. As in \cite{bcov2} (henceforth 
referred to simply as BCOV), the holomorphic anomaly is a statement 
about the anti-holomorphic derivative $\delbar \F gh$. While naively 
zero, it turns out that $\delbar \F gh$ receives a contribution from, 
and only from, the boundary $\del\calm gh$ of $\calm gh$, where the term 
``boundary'' here refers in the complex sense to the parts of $\calm gh$ 
that have been added to the space of actual Riemann surfaces to make a 
compact $\calm gh$. In other words, the boundary term arises from the 
singularities or contact terms that appear in the integrand of
$\F gh$ when the Riemann surface degenerates. The key to the 
holomorphic anomaly is that the boundary term itself is not a 
holomorphic function over $M$. 

For the closed string, $(g,h)=(g,0)$, the Riemann surface can degenerate
in one of two ways. It can either split into two Riemann surfaces of 
lower genus $g_1$ and $g_2$, or a handle can pinch, leaving a Riemann 
surface of genus $g-1$. The holomorphic anomaly equation for $\Fc g\equiv
\F g0$ then takes the form (for $g>1$; for $g\le 1$, see below)
\begin{equation}
\eqlabel{master}
\del_\ib \Fc g = \frac 12 C_{\ib \jb\kb} \ee^{2K} G^{\jb j} G^{\kb k}
\Bigl(\sum_{g_1+g_2=g} \Fc {g_1}_j \Fc {g_2}_k + \Fc {g-1}_{jk} \Bigr)
\end{equation}
Here, $\Fc g$ with subscripts, $i$, $j$, \etc, refer to amplitudes with
insertions, \ie, derivatives of $\Fc g$ in holomorphic directions on $M$,
and $C_{ijk}\equiv \Fc{0}_{ijk}$ is the three-point function on the sphere 
(Yukawa coupling), which is holomorphic (whence $C_{\ib\jb\kb}\equiv
\overline{C_{ijk}}$ is anti-holomorphic). (See the next section for the 
precise explanation of all 
the symbols.) What is important to recognize here is that the sum involves
only $g_1,g_2<g$, so that eq.\ \eqref{master} really is a recursive relation 
for the topological amplitudes genus by genus. The underlying reason is that 
the amplitude on the sphere with less than three insertions vanishes.

For the open string, $h\neq 0$, we first of all have to choose boundary 
conditions, in other words we have to specify a D-brane configuration.
Then, we can write an extension of \eqref{master} to the open string at 
the conjuncture of the following four fundamental facts.

\begin{list}{}{\setlength{\leftmargin}{.3cm}\setlength{\rightmargin}{0cm}%
\setlength{\labelwidth}{.2cm}\setlength{\itemindent}{.2cm}%
\setlength{\labelsep}{.2cm}%
\setlength{\topsep}{.2cm}\setlength{\itemsep}{.2cm}\setlength{\parsep}{0cm}}

\item[(F1)] For generic values of the bulk moduli, the topological amplitudes
do not depend on any continuous open string moduli. To justify this, we note 
that the topological disk amplitude, $\F 01$, is interpreted physically as 
the 4d superpotential on the brane worldvolume. Brane moduli correspond to 
flat directions of this superpotential, so $\F 01$ cannot depend on them. 
Since the holomorphic anomaly ultimately reduces everything to tree-level
information, we do not expect any $\F gh$ to depend on open string moduli.
Note, however, that we are {\it not} claiming that open string moduli are 
generically absent, or otherwise uninteresting, just that we can 
ignore them for our present purposes.

\item[(F2)] The topological charge of the D-brane configuration under 
consideration vanishes. This can be traced back to the statement that 
the physical quantity we are computing at tree-level is the tension of BPS
domainwalls between various brane vacua, in other words superpotential
differences, and not directly the value of the superpotential itself.
Note that the topological charges are carried by the ground states 
corresponding to marginal directions of the ``other'' topological
string (for B-branes---the A-model, for A-branes---the B-model), which
are BRST trivial and should decouple. We therefore view this restriction 
as a kind of topological tadpole cancellation condition.

\item[(F3)] The disk amplitude with two bulk insertions is the analogue
of the closed string Yukawa coupling. This is in fact easy to see.
The Yukawa coupling is so central data because it is the first non-vanishing 
amplitude at tree-level (the sphere 0, 1, and 2-point functions vanish),
and all higher-point functions on the sphere can be computed from it by 
simply taking derivatives. (Although, taking derivatives requires information 
not contained in the Yukawa coupling itself.) For open strings, the naively
simplest quantities to consider are the disk amplitude with three boundary
insertions or with one bulk and one boundary insertion. Both precisely
cancel the ghost-number anomaly on the disk. However, given (F1) above, we 
generically do not have any non-trivial operators to insert on the boundary,
and then the first non-vanishing quantity is indeed the disk two-point 
function. Since one of the insertions then has to be (half-)integrated,
we find that the disk two-point function is itself not holomorphic, in clear
distinction to the closed string, where the Yukawa coupling always remains
holomorphic. The holomorphic anomaly equation for the disk two-point function
can be viewed as the open string analogue of special geometry.

\item[(F4)] For $2g+h-2>0$, the holomorphic anomaly receives no contribution 
from factorization in the open string channel. This can be understood as a 
consequence of the rule that only moduli fields could contribute in such
factorizations (\cf, \eqref{master}), and those are excluded by (F1) above.
As a consequence, the only degenerations which contribute new terms on the 
RHS of \eqref{master} are those where the length of a boundary component 
shrinks to zero size. The only exception to this rule occurs for $(g,h)=(0,2)$, 
\ie, the annulus amplitude. This exception is a direct counterpart of the 
anomaly of the torus amplitude $(g,h)=(1,0)$.

\end{list}

(F4) above determines the general structure of the extended holomorphic
anomaly equation, whereas (F3) gives the basic connection to geometric data
at tree-level. To immediately\footnote{and very imprecisely; the patient reader
is urged to skip the next paragraph or two.} give away the punchline of this 
identification, let us first recall the computation of the Yukawa coupling in 
the B-model. If $\Omega(z)\in H^{3,0}(Y)$ denotes the holomorphic three-form 
as a function of the complex structure moduli of the Calabi-Yau manifold $Y$, 
then the Yukawa coupling (which is mirror to the instanton corrected triple 
intersection in the A-model) can be computed as
\begin{equation}
C_{ijk} = -\langle \Omega, \nabla_i\nabla_j\nabla_k \Omega\rangle
= -\langle \Omega,\del_i\del_j\del_k\Omega\rangle
\end{equation}
where $\nabla$ is the Gauss-Manin connection, and $\langle\cdot,
\cdot\rangle=\int_Y \cdot\wedge\cdot$ the symplectic pairing on $H^3(Y)$.
The equality with ordinary derivatives is a consequence of {\it Griffiths 
transversality},
\begin{equation}
\eqlabel{transverse}
\langle \Omega,\nabla\nabla\Omega\rangle=\langle\Omega,\nabla\Omega\rangle=0 \,.
\end{equation}

Now consider open strings. It will follow from (F1), (F2) above, and is further
explained in \cite{mowa} that the invariant holomorphic data that characterizes
a topological D-brane at tree-level is a {\it Poincar\'e normal function}, in the sense
introduced by Griffiths in his early work \cite{griffiths1} on higher-dimensional
Hodge theory. Very briefly, a normal function, $\nu$, is defined by a three-chain 
$\Gamma\subset Y$ whose boundary is a holomorphic curve. Such a three-chain does 
not quite specify an element in $H^3(Y)$. But because the boundary is 
holomorphic, integration against $\Gamma$ in any case gives a well-defined pairing 
with cohomology classes in $H^{3,0}(Y)$ and $H^{2,1}(Y)$. Physically, we identify
\begin{equation}
\eqlabel{given}
\calt = \langle\Omega,\nu\rangle = \int_\Gamma \Omega
\end{equation}
with the domainwall tension. The hallmark of a normal function is its
own version of {\it Griffiths transversality} \cite{griffiths1}
\begin{equation}
\langle\Omega, \nabla\nu\rangle = 0
\end{equation}
obviously the analogue of \eqref{transverse}. All the local information about
$\nu$ is then contained in the {\it Griffiths' infinitesimal invariant} 
\cite{griffiths2,voisin,green}, which we identify with the disk two-point 
function,
\begin{equation}
\eqlabel{strictly}
\F 01_{ij} = \Delta_{ij} = \langle\Omega,\nabla_i\nabla_j\nu\rangle \,.
\end{equation}
Let us pause. Mathematically, the infinitesimal invariant is not known as
a symmetric tensor defined by \eqref{strictly}, but only as a certain Koszul
cohomology class whose {\it representative} depends on a lift of $\nu$ to 
$H^3(Y;\complex)$. But, given \eqref{given} and its physical interpretation, 
there is a {\it preferred lift} given by declaring $\nu$ to be real, \ie, $\nu\in 
H^3(Y;\reals)\subset H^3(Y;\complex)$. Conspicuously, reality is not 
compatible with holomorphic dependence on the parameters. But {\it this is 
precisely the holomorphic anomaly of the disk two-point function expected
from (F3) above!} And with this point driven home, we can put off 
all further explanations to the next section.

\subsection{Examples}

The example on which we will test the extended holomorphic anomaly
equation is in the A-model given by the quintic hypersurface $X\subset
\projective^4$, where we wrap a D-brane on the real locus $L\subset X$ 
inside of it. The B-model mirror of $X$ is well-known as the mirror 
quintic, and the D-brane is best described by a certain matrix 
factorization of the corresponding Landau-Ginzburg potential. The 
holomorphic tree-level data for this brane configuration was determined 
in \cite{open}, further explanations appear in \cite{mowa}.

The main point of \cite{open} was that the domainwall tension \eqref{given} 
for the real quintic satisfies an extended or inhomogeneous version of
the Picard-Fuchs differential equation governing periods of the 
mirror quintic,
\begin{equation}
\call \calt(z) = c \sqrt{z},
\end{equation}
where $\call$ is the Picard-Fuchs operator. The value of the
constant $c$ was determined in \cite{open} from monodromy 
considerations, and checked against explicit computations in 
the A- and B-model in \cite{psw} and \cite{mowa}, respectively.
This constant will prove of crucial importance 
for the present purposes, as it allows us to identify the correct 
real lift of the corresponding normal function.

We can then plug this tree-level data into the holomorphic
anomaly equations. We will here solve these equations for small
$(g,h)$ in a rather pedestrian fashion, compared with the best currently 
available technology. In particular, we will not attempt to formulate
a polynomial solution as done by Yamaguchi and Yau in \cite{yayau}
for the closed string. But we note that the structure of the equations 
strongly suggests that this is straightforwardly possible.

We then also need to fix the holomorphic ambiguity, which is one of
the central problems in using the holomorphic anomaly equations to
solve the topological string. In the closed string case, constraints
on the holomorphic ambiguity arise from the physical expectations
on the expansion of the $\Fc{g}$ around the special or singular loci 
in the moduli space. This was convincingly pushed to very high 
order in recent work by Huang, Klemm and Quackenbush \cite{hkq}. For 
the quintic, the three special points are large volume, conifold, 
and Gepner (or orbifold) point. The expansion of topological string 
amplitudes around large volume has been shown to capture the BPS content 
of an M-theory compactification, and as a consequence to satisfy a 
certain integrality property known after Gopakumar and Vafa \cite{gova}. 
Around the conifold, the $\Fc g$ are generally singular, but with a 
singularity structure determined by the appearance of precisely one 
massless BPS particle in the corresponding string compactification. 
Finally, the regularity of $\Fc {g}$ around the Gepner point also 
imposes additional constraints.
 
For the open string, we also have an integrality conjecture at large 
volume proposed by Ooguri and Vafa \cite{oova}, and refined by Labastida, 
Mari\~no, and Vafa \cite{lmv}. According to this conjecture, the open 
topological string amplitudes count degeneracies of BPS states 
(domainwalls) in the 2-dimensional worldvolume theory of a brane 
partially wrapped on the Calabi-Yau. We also expect some sort of 
singularity structure at the conifold. The main novelty for us is that 
the Gepner point is {\it not a regular point} once open strings are 
included. As pointed out in \cite{bdlr,howa}, the D-brane under consideration 
exhibits an extra massless open string in its cohomology precisely at the 
Gepner point. As a consequence \cite{open}, a tensionless domainwall 
appears in the BPS spectrum, and this is expected to leave an imprint on 
the topological string amplitudes $\F gh$ for $h>0$. (The attentive reader 
will see this a qualification of the statement (F1) above.) We cannot at 
present describe the singularity structure in general. Nevertheless, with 
certain (likely too optimistic) assumptions, we will be able to fix the 
holomorphic ambiguity for the topological amplitudes $\F 02$, $\F 03$ and 
$\F 11$. In particular, we will find an integral structure around large 
volume, as would be predicted from the existence of BPS invariants.

\subsection{Conclusions}

From what we have said in this introduction, it is clear that this
work is a direct generalization of BCOV to the open string sector,
and has bypassed many of the intervening developments on the structure
of the holomorphic anomaly, techniques for solving it, as well as
relations to target space physics. It will be very interesting to
revisit these various connections.

\enlargethispage{.6cm}

\section{Extended Holomorphic Anomaly}

We will begin by refreshing the main ideas, results, and derivations from 
BCOV \cite{bcov2}. The main purpose of subsections \ref{Neq2twisted}, 
\ref{vacgeo1}, \ref{cshan} is to establish notation and collect some 
useful formulas. The interlaced subsection \ref{first} contains a
few basic points about D-branes in the topological string, emphasizing
the deformation and obstruction theory that is needed to understand
fact (F1) from the introduction. In subsection \ref{more}, we 
describe how D-branes fit into the geometry of the vacuum bundle.
After these preparations, we are then ready to derive the
holomorphic anomaly equation for the open string, for the disk 
in subsection \ref{infinvhan}, and for higher topologies in 
\ref{openhan}. In subsection \ref{oneloophan},
we discuss the special status of the holomorphic anomaly at one
loop in both closed and open string. In subsection \ref{insertions},
we write down the holomorphic anomaly for amplitudes with chiral
insertions. Finally, in subsection \ref{hansolutions}, we will
develop the basic techniques for solving the extended holomorphic
anomaly equation, in parallel to the method of BCOV.

\subsection{Twisted \texorpdfstring{$\caln=2$}{N=2} theories}
\label{Neq2twisted}

The starting point for the definition of a topological string theory is a
2-dimensional conformal field theory with $\caln=(2,2)$ worldsheet
supersymmetry. Such theories have a total of four real supercharges,
two of holomorphic origin (on the worldsheet), $G^\pm$, and two 
anti-holomorphic ones, $\Gb^\pm$. Here, the superscript indicates the 
charge under the $U(1)$ R-symmetries, $J$, $\Jb$. We will for simplicity
directly assume that the central charge of the theory is $\hat c=3$, and
that all $U(1)$ charges are integer. The holomorphic supercharges satisfy 
the algebra
\begin{equation}
\eqlabel{algebra}
(G^\pm)^2=0 \,, \qquad \{G^+,G^-\} = 2 L_0 \,, \qquad
[G^\pm ,L_0] = 0
\end{equation}
where $L_0$ denotes the zero mode of the holomorphic stress-tensor. The 
anti-holomorphic version of the algebra is similar.

Among the important operators in an $\caln=(2,2)$ SCFT are the chiral primary
operators, which are defined from the cohomology of the supercharges. The 
chiral operators form a ring and are in one-to-one correspondence with the 
supersymmetric ground states of the theory \cite{lvw}. There are in fact four 
different rings that can be constructed, depending on the combination of 
holomorphic/anti-holomorphic supercharges (see table \ref{four}). The $U(1)$
R-symmetries provide the rings with two gradings, which we will denote by $q$ 
and $\qb$. Fields which are chiral primary on the holomorphic side have
$0\le q\le \hat c$, while the anti-chiral ones have $0\ge q\ge\hat c$, and
similarly for the anti-holomorphic side. The $U(1)$ charge of the corresponding
RR ground states, which can be reached from each of the chiral rings by spectral 
flow, lie between $-\hat c/2$ and $\hat c/2$.

Two discrete symmetries of an $\caln=2$ SCFT are of particular importance
for the topological theory. The first one is simply CPT invariance on the
worldsheet, which identifies the $(c,c)$ ring with the $(a,a)$ ring, and
the $(c,a)$ ring with the $(a,c)$ ring. The other symmetry is just as
obvious from the algebra \eqref{algebra}, but more subtle in its consequences. 
It is the mirror automorphism which exchanges the $(c,c)$ with the $(c,a)$
ring and the $(a,a)$ with the $(a,c)$ ring. For most of the discussion in
topological strings, only two of the rings are relevant at the same time. 
To be specific, we will concentrate on the topological B-model, and its 
conjugate counterpart, the anti-topological B-model. We will sometimes
refer to the A-model as ``the other model''.

\begin{table}[ht]
\begin{center}
\begin{tabular}{|l|c|c|c|}
\hline
model & chiral ring &  BRST charges & anti-ghosts \\\hline
topological A-model & $(c,a)$ & $G^+ , \Gb^-$ & $G^-, \Gb^+$ \\
anti-topological A-model & $(a,c)$ & $G^-, \Gb^+$ & $G^+ , \Gb^-$ \\
topological B-model & $(c,c)$  & $G^+ , \Gb^+$ & $G^- , \Gb^-$ \\
anti-topological B-model & $(a,a)$ & $G^- , \Gb^-$ & $G^+ , \Gb^+$ 
\\\hline
\end{tabular}
\end{center}
\caption{Four different topological models can be constructed from
any $\caln=(2,2)$ super-conformal field theory.}
\label{four}
\end{table}

Part of the interest of the chiral rings arises from the fact that the
subset of fields of charge $(q,\qb)=(1,1)$ parametrize deformations of the
SCFT. Given an infinitesimal chiral primary $\phi$ with those charges, we 
can deform the theory by adding to the action,
\begin{equation}
\eqlabel{deform}
\delta S = \int d^2zd^2\theta\phi + \int d^2zd^2\bar\theta\phib
= \int \phi^{(2)} + \int \phib^{(2)}
\end{equation}
where $\phib$ is the anti-chiral field conjugate to $\phi$.
Also, $\phi^{(2)} = dzd\zb\{G^-,[\Gb^-,\phi]\}$ will be the two-form 
descendant of $\phi$.

The next step in the construction is the topological twist of 
the SCFT into a {\it topological field theory}. The twist 
amounts to redefining the worldsheet stress tensor $T\to T\pm \del J$,
or equivalently to couple the $U(1)$ R-symmetry current to a background
connection which is equal to the spin connection on the worldsheet.
Just as there are four chiral rings, there are also four different
topological twists. After the topological twist, half of the supercharges
become scalar, the other half one-forms, and the algebra \eqref{algebra} 
coincides with the algebra satisfied by the {\it BRST operator and anti-ghost
in the critical bosonic string}. The $U(1)$ charges are identified with 
the ghost numbers. More specifically, for the topological B-model, one 
identifies
\begin{equation}
\eqlabel{identify}
2 Q_{{\it BRST}} \leftrightarrow G^+\,,
\qquad b_0 \leftrightarrow G^-\,,
\qquad bc \leftrightarrow J
\end{equation}
While formally similar, there are three important differences to the
bosonic string. Firstly, there is no ghost field, or more precisely, 
ghost and matter fields are not decoupled from one another. Secondly,
we have a finite-dimensional BRST cohomology. The Hilbert space of 
closed string physical states decomposes according to the grading of 
the chiral ring by the two $U(1)$ charges,
\begin{equation}
\eqlabel{hilbert}
\calh_{\rm closed} = \bigoplus_{q,\qb=0}^3 \calh^{q,\qb}
\end{equation}
Finally, the cohomology of the anti-ghost is non-trivial. This is obvious
from the definition since by the identification \eqref{identify}, the 
anti-ghost cohomology is simply the anti-chiral ring, which is isomorphic 
to the chiral ring by worldsheet CPT. This has profound consequences,
among others the holomorphic anomaly. But at first, these modifications 
appear minor, and the structure obtained by the topological twist of a 
unitary $\caln=2$ SCFT is sufficient to define a measure on the moduli 
space of Riemann surfaces just as in the bosonic string. This is the 
{\it topological string}.

\subsection{Geometry of the vacuum bundle}
\label{vacgeo1}

The most interesting aspects of the structure of $\caln=(2,2)$ SCFTs and their
topological twists are revealed when one considers them in families. It was 
already mentioned above that one can parametrize the infinitesimal deformations
of the topological B-model by the chiral fields of charge $(q,\qb)=(1,1)$.
These deformations are in fact all unobstructed and span a complex
manifold, $M$, of dimension $n=\dim \calh^{1,1}$. We will now continue to 
follow BCOV and concentrate on the subring of the chiral ring generated 
by the marginal fields. If $(\phi_i)$ for $i=1,\ldots,n$ is a basis of
marginal fields, a basis for the subring they generate is given by 
$(\phi_0,\phi_i,\phi^i,\phi^0)$. Here, $\phi_0$ is the identity operator
of charge $(q,\qb)=(0,0)$, and $\phi^i$ are the charge $(2,2)$ fields 
which are dual to $\phi_i$ with respect to the topological metric
\begin{equation}
\eqlabel{metric1}
\eta(\phi_i,\phi^j) = \langle\phi_i \phi^j\rangle_{0} = \delta_i^j
\end{equation}
where $\langle\,\cdot\,\rangle_{g=0}$ denotes the correlation function of the 
topological field theory on the sphere. Finally, $\phi^0$ is the top element 
in the chiral ring, of charge $(3,3)$, and satisfies
\begin{equation}
\eqlabel{metric2}
\eta(\phi_0,\phi^0) = \langle\phi_0\phi^0\rangle_0 = 1
\end{equation} 
The ring structure is encoded in the three-point function on the sphere, 
also known as the Yukawa coupling,
\begin{equation}
\eqlabel{yukawa}
C_{ijk} = \langle \phi_i\phi_j\phi_k\rangle_0
\end{equation}
Namely,
\begin{equation}
\phi_i\phi_0 = \phi_i\,,\qquad
\phi_i\phi_j = C_{ijk} \phi^k\,,\qquad
\phi_i\phi^j = \delta_i^j\phi^0\,,\qquad
\phi_i\phi^0 = 0
\end{equation}
Note that by topological invariance, it does not matter where on the sphere
we insert the operators in either the definition of the metric or the Yukawa
coupling.

As we move around in the moduli space $M$, the space of vacua with $q=\qb$
generated by the moduli fields fit together into a holomorphic vector bundle 
known as the {\it vacuum bundle} $\calv\to M$. At any point $m\in M$, we can 
decompose
\begin{equation}
\eqlabel{decompose}
\calv_m = \calh^{0,0}\oplus \calh^{1,1}\oplus\calh^{2,2}\oplus\calh^{3,3}
\end{equation}
As we have done before, we will use the operator-state correspondence to
identify the basis of the chiral ring with a basis for $\calv_m$. Namely,
we let $e_0\in \calh^{0,0}$ be the unique-up-to-scale ground state of 
charge $(0,0)$, and then obtain a basis of $\calv_m$ by
\begin{equation}
\eqlabel{rrbasis}
e_i = \phi_i e_0\,,\qquad e^i = \phi^i e_0\,,\qquad e^0 = \phi^0 e_0\,.
\end{equation}
The topological metric \eqref{metric1}, \eqref{metric2} is then a metric 
on the vacuum bundle,\footnote{'$\langle$' and '$\rangle$' are somewhat
over-used in this context. Our conventions are that $\langle\cdot
\rangle_{(g,h)}$ denotes worldsheet correlators, $\langle\cdot|\cdot\rangle$ 
the symmetric topological metric, and $\langle\cdot,\cdot\rangle$ the 
symplectic pairing, which is anti-symmetric. We will try to be consistent.}
\begin{equation}
\eta(e_a,e^b) = \langle e_a|e^b\rangle = \delta_{a}^b\,,\qquad
\text{for $a,b=0,\ldots n$}
\end{equation}
 
Another essential ingredient for the holomorphic anomaly is the existence
of another metric on $\calv$ besides the topological one, \eqref{metric1}, 
\eqref{metric2}. This metric is known as the $tt^*$-metric and its
definition depends in an essential way on the unitarity of the underlying
$\caln=2$ SCFT. If $\Theta$ is the CPT operator acting on the ground states,
we define the $tt^*$-metric by \cite{ceva}
\begin{equation}
\eqlabel{ttstar}
g_{a \bb} = g(e_b,e_a) = \langle \Theta b| a \rangle
\end{equation}
(If one wants to define this inner product by a path integral as in 
\eqref{metric1}, one has to be more careful about where one inserts the
operators.) Using the $tt^*$-metric, one can define a new basis for
the charge 2 and 3 subbundles of $\calv$, via
\begin{equation}
\eqlabel{newbasis}
e_\ib = e^k g_{k\ib} \qquad
e_\0b = e^0 g_{0 \0b } 
\end{equation}

The set of data discussed above satisfies a number of relations, known as 
the $tt^*$-equations \cite{ceva}, and which specialize to special geometry 
for $\hat c=3$. Let us write out these equations for future reference.

First of all, the $tt^*$-metric induces a connection on the bundle
$\calv$. This connection is simply the unique one compatible with the 
metric and the holomorphic structure on $\calv$. With respect to the 
basis $(e_a) = (e_0,e_i,e_\ib,e_\0b)$, the connection matrix of the 
$tt^*$-connection, $D_i (e_a) = (A_i)_a^b e_b$, is given by the usual 
formula, $A_i = g^{-1}\del_i g$, or explicitly
\begin{equation}
\eqlabel{ttstarcon}
A_i = \begin{pmatrix}
g^{\0b 0}\del_ig_{0\0b}&&&\\
& g^{\jb l}\del_i g_{m \jb} &  & \\
 &  &  0 &  \\
 &  &   & 0 
\end{pmatrix}\,,
\qquad
A_\ib = 
\begin{pmatrix}
0 &  &  &  \\
 & 0& & & \\
 & & g^{\lb k}\del_\ib g_{k\mb} &\\
& & & g^{\0b 0}\del_\ib g_{0\0b}
\end{pmatrix} 
\end{equation}
Furthermore, the vacuum bundle carries an action of the chiral fields, as
we have already used in the definition of the basis \eqref{rrbasis}.
In matrix representation, multiplication by the chiral fields $\phi_i$,
$\phi_\ib$ is explicitly,
\begin{equation}
\eqlabel{struccon}
C_i = \begin{pmatrix}
0&0&0&0\\
\delta_i^l & 0 & 0 & 0\\
0 & C_{im}^{\;\;\;\;\lb} & 0 & 0\\
0&0&G_{i\mb}&0
\end{pmatrix}\,,
\qquad
C_\jb =
\begin{pmatrix}
0&G_{\jb m}&0&0\\
0&0&C_{\jb\mb}^{\;\;\;\;l} &0\\
0&0&0&\delta_{\jb}^\lb\\
0&0&0&0
\end{pmatrix}
\end{equation}
where $C_{im}^{\;\;\;\;\lb}:=C_{iml} g^{\lb l}$. The $tt^*$-connection 
and the multiplication by the chiral ring satisfy the so-called 
$tt^*$-equations,
\begin{gather}
[D_i,D_j]=[\Db_\ib,\Db_\jb]=[D_i,\Cb_\jb] = [\Db_\ib,C_j]=0\nonumber \\
\eqlabel{ttstareq}
[D_i,C_j]=[D_j,C_i]\qquad [\Db_\ib,\Cb_\jb]=[\Db_\jb,\Cb_\ib]\\\nonumber
[D_i,\Db_\jb] = -[C_i,\Cb_\jb]
\end{gather}
These equations are equivalent to the flatness of the one-parameter family
of ``improved'' connections on $\calv$, often referred to as the Gauss-Manin 
connection,
\begin{equation}
\nabla_i = D_i - \alpha C_i \qquad \nabla_\ib = \Db_\ib - \alpha^{-1} \Cb_\ib
\end{equation}
The value of the parameter $\alpha$ in identifying $\nabla$ with
the geometric Gauss-Manin connection has to do with the existence of a real
structure on the vacuum bundle, which is a point to which we shall return
below in subsection \ref{more}.

For $\hat c=3$, the $tt^*$-equations can be formulated more intrinsically
in terms of the geometry of the moduli space $M$ itself (as opposed to 
the vacuum bundle over it). As we have noted, there is an identification
between the charge $(1,1)$ subbundle of $\calv$ and the tangent bundle
of $M$. The identification involves the charge $(0,0)$ space $\calh^{0,0}$,
which forms a holomorphic line bundle $\call$ over $M$. Namely,
\begin{equation}
\eqlabel{ident}
\calh^{1,1} \cong \call\otimes TM
\end{equation}
The charge $(2,2)$ and $(3,3)$ subbundles can be identified with the duals
of $\call\otimes TM$ and $\call$, respectively, by using the topological metric, 
or with their hermitian conjugates by using the $tt^*$-metric.

A metric on $M$, known as the Zamolodchikov metric, can be defined by
restricting the $tt^*$-metric to $\calh^{(1,1)}$ and using the identification
\eqref{ident}
\begin{equation}
G_{i\jb} = \frac{g_{i\jb}}{g_{0\0b}}
\end{equation}
It follows from the $tt^*$-equations that the Zamolodchikov metric is a 
K\"ahler metric on $M$, with K\"ahler potential $K=-\log g_{0\0b}$.
Namely,
\begin{equation}
\eqlabel{kahler}
G_{i\jb} = \del_i\del_\jb K
\end{equation}
Moreover, the Yukawa coupling is a symmetric rank 3 tensor with values in 
$\call^{-2}$ which is holomorphic, and whose covariant derivative is symmetric 
in all four indices,
\begin{equation}
\eqlabel{symm}
\del_\lb C_{ij,k} = 0 \,,\qquad D_{i}C_{jkl} = D_j C_{ikl}\,.
\end{equation}
Here, by abuse of notation, we are using $D$ to denote the natural connection 
on ${\rm Sym}^3T^*M\otimes\call^{-2}$, given by the sum of the Zamolodchikov 
connection (the metric connection for $G_{i\jb}$) and the K\"ahler connection 
on $\call$. (This is consistent with $D$ being the $tt^*$-connection on 
$\call\otimes TM$.) Finally, the curvature of the Zamolodchikov metric is 
(in its representation on the tangent bundle, with basis $\phi_i$)
\begin{equation}
\eqlabel{riemann}
(R_{i\jb})^k_{\;l} = [D_i,D_\jb]^k_{\;l} = C_{ilm} C_{\jb\mb\kb} \ee^{2K}
G^{\mb m} G^{\kb k} - \delta_l^k G_{i\jb} - \delta_i^k G_{l\jb}
\end{equation}
Many formulas compactify if we agree to raise and lower indices with
the $tt^*$-metric, \eg, 
\begin{equation}
C_\ib^{jk}:=C_{\ib\jb\kb} g^{\jb j}g^{\kb k}
= C_{\ib\jb\kb}\ee^{2K} G^{\jb j}G^{\kb k}
\end{equation}
The conditions \eqref{kahler}, \eqref{symm}, \eqref{riemann} are precisely 
equivalent to what is known as a {\it special K\"ahler structure} on the 
moduli space $M$.

\subsection{First comments on boundaries}
\label{first}

Boundary conditions in $\caln=2$ theories have been studied in many works
over the years since BCOV. We will here recall some standard and some 
possibly less-well appreciated facts (for background material see 
\cite{horietal}), and then quickly move to the generalization of the 
structures described in the previous subsection, which is our main 
interest in this paper.

Recall that starting from an $\caln=(2,2)$ CFT, we can construct two
different topological theories, the A-model and the B-model, depending
on which combination of left and right-moving supercharges becomes the 
BRST operator. Since defining D-branes involves choosing boundary 
conditions between left and right-movers, this means we can also 
consider two kinds of D-branes, called A-branes and B-branes, 
respectively \cite{oooy}. For example, the boundary conditions for 
B-branes are,
\begin{equation}
\eqlabel{boundary}
\bigl(G^-+\Gb^-\bigr)|_{\del\Sigma} = 0\,, \qquad
\bigl(G^++\Gb^+\bigr)|_{\del\Sigma}=0\,, \qquad
\bigl(J - \Jb\bigr)_{\del\Sigma}= 0
\end{equation}
B-type boundary conditions are compatible with B-type topological
twist in the sense that we can define topological string amplitudes
with background D-branes that are BRST invariant.

Somewhat oddly at first, it also makes sense to consider A-type
boundary conditions when one is in the B-model, and B-type boundary 
conditions in the A-model. To see the relevance of the branes from the 
other model, consider the overlaps of some B-brane $B$, boundary state
$|B\rangle$, with the supersymmetric (Ramond-Ramond) ground states, 
which compute the topological charges of D-branes modulo torsion,
\begin{equation}
\langle {\it rgs} | B \rangle 
\end{equation}
Using \eqref{boundary}, it is not hard to show that this vanishes
unless $q_{\it rgs}+\qb_{\it rgs}=0$, namely the vector R-charge
of the ground state has to vanish. These ground states {\it do not}
correspond to the marginal $(c,c)$-fields if one uses spectral flow
to the B-model, but rather to the marginal $(q,\qb)=(1,-1)$ fields
from the $(c,a)$ ring in the topological A-model. Somewhat informally,
one can think that the topological charges of B-branes are carried 
by the A-model (and vice-versa).\footnote{This interaction of A-
and B-model has also become familiar in recent years in the context of 
stability conditions \cite{dougstab,bridgeland,thomasyau}. For Lagrangian 
(A-)branes, they involve the holomorphic three-form, which is a B-model 
quantity, whereas for B-branes, stability depends on the (complexified, 
quantum corrected) K\"ahler form.} Conversely, it means that if we want 
to use D-branes to probe the structure of the vacuum bundle (see subsection 
\ref{more}), we have to take them from the other model.

For now, let us discuss aspects of B-branes in the B-model. For
boundary conditions preserving $\caln=2$ supersymmetry, the essentials
of the discussion on chiral rings, their relation to (open string)
supersymmetric ground states, \etc, remain unchanged. The main
difference is that we have only one R-charge to label the states
and fields,
\begin{equation}
\eqlabel{openhilbert}
\calh_{\rm open} = \oplus_{p=0}^3 \calh^p 
\end{equation}
The relation between bulk and boundary R-charges is
\begin{equation}
\eqlabel{relation}
p = q + \qb
\end{equation}
We will generically denote elements of the boundary chiral ring
by $\psi_a$, or $\psi_i$ for the marginal fields with $p=1$. We
will write $u^i$ for the corresponding worldsheet couplings. In 
distinction to the closed string case, these deformation are not 
always unobstructed. Rather, there can be a higher order superpotential
$\calw=\calw(u)$, whose critical points as a function of the $u^i$ 
determine the supersymmetric vacua of the D-brane theory. 

\begin{figure}
\begin{center}
\psfrag{phi}{$\phi$}
\psfrag{psi}{$\psi$}
\psfrag{psi1}{$\psi_1$}
\psfrag{psi2}{$\psi_2$}
\psfrag{psi3}{$\psi_3$}
\psfrag{psin}{$\psi_{n+1}$}
\epsfig{height=4cm,file=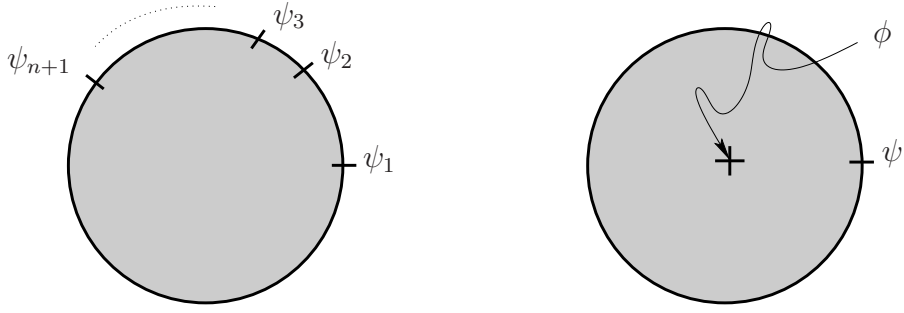}
\caption{Left: The disk amplitude with $n+1$ boundary insertions captures 
the higher $A_\infty$ products \eqref{mn}. Right: The disk amplitude with 
one bulk and one boundary insertion encodes the bulk to boundary 
obstruction map \eqref{obmap}, and will also define the infinitesimal 
Abel-Jacobi map \eqref{infaj}.}
\label{obstruct}
\end{center}
\end{figure}

In the categorical approach to D-branes on Calabi-Yau manifolds, 
the obstructions and the superpotential are succinctly encoded in 
a so-called $A_\infty$-structure on the triangulated D-brane category. 
(The study of $A_\infty$-structures to which the present discussion is
closest in spirit appears in \cite{hll}. See \cite{herbst} for
an extension to higher worldsheet topologies. The homological background 
for D-brane physics is explained in many works, see for instance
\cite{sharpe,dougcat,calinreview,sharpereview}.)
More precisely, if $B, B'$ are two B-branes, the topological Hilbert 
spaces of $B$--$B'$-strings are identified as ${\rm Ext}$-groups 
between the objects in the category
\begin{equation}
\calh^p_{B-B'} \cong {\rm Ext}^p(B,B')
\end{equation}
Infinitesimal deformations of a brane $B$ correspond to 
${\rm Ext}^1(B,B)$, obstruction spaces to ${\rm Ext}^2(B,B)$. Furthermore,
there is a collection of higher-order obstruction maps,
\begin{equation}
\eqlabel{mn}
m_n : \bigl({\rm Ext}^1(B,B)\bigr)^{\otimes n} \to {\rm Ext}^2(B,B)
\,,\qquad (n\ge 2)
\end{equation}
which by using the topological open string metric can be identified
with the $n+1$-point function on the disk, see Fig.\ \ref{obstruct}.
The worldvolume superpotential can then be defined as \cite{calin}
\begin{equation}
\eqlabel{supo}
\calw_B(u) = \sum_{n=2}^\infty \frac 1{n+1}\langle u|m_n(u^{\otimes n})\rangle\,.
\end{equation}
Its critical points correspond precisely to the locus where the 
higher-order obstruction maps vanish.

Not only are open string deformations often obstructed by the non-vanishing
of higher products on the ${\rm Ext}$-groups, but the very presence of
background D-branes can sometimes obstruct the closed string deformations. 
Since this will be important later on, let us give a brief worldsheet 
derivation of this fact.

A basic feature of F-terms in supersymmetric theories is that they are
supersymmetric only up to a total derivative. In spacetimes or on 
worldsheets with boundary therefore, this can lead to a non-vanishing
boundary term in the supersymmetry variation. Consider in our framework
the deformation of the bulk action as in \eqref{deform}. By using the
topological descent relations
\begin{equation}
\{G^+,\phi^{(2)}\} = \del\phi^{(1)}\,,\qquad 
\{\Gb^+,\phi^{(2)}\} = \delbar\phi^{(1)}
\end{equation}
we see that its supersymmetry variation on a worldsheet, $\Sigma$, with 
non-empty boundary produces a boundary term,
\begin{equation}
\eqlabel{warner}
\delta_Q (\delta S) =  \int_\Sigma \{G^++\Gb^+,\phi^{(2)}\} =
\int_\Sigma (\del+\delbar)\phi^{(1)} = 
\int_\Sigma d\phi^{(1)} 
= \int_{\del\Sigma} \phi^{(1)} \quad \bigl(+ {\it c.c.}\bigr)
\end{equation}
In the context of topological field theories, the boundary term on the 
RHS of \eqref{warner} has come to be known as ``the Warner problem'' 
\cite{warner}.

By using \eqref{relation}, the boundary $U(1)$ charge of 
$\phi^{(1)}|_{\del\Sigma}$ is $p=2$, and since it is still BRST closed,
we obtain a well-defined open string chiral field from $\calh^2_{\rm open}$.
This defines the {\it bulk-to-boundary obstruction map}
\begin{equation}
\eqlabel{obmap}
m_0 : TM \cong \calh^{1,1}_{\rm closed} \to {\rm Ext}^2(B,B) \cong 
\calh^2_{\rm open}
\end{equation}
which as indicated fits as a zeroth order product into the 
$A_\infty$-framework. (An $m_1$ can be identified as the BRST
operator itself.) Diagrammatically, $m_0$ composed with the open 
string topological metric can be defined as the disk correlator with 
one bulk and one boundary insertion, see Fig.\ \ref{obstruct}.

The further fate of the Warner problem depends on whether $m_0(\phi)$ 
vanishes in cohomology or not. If it is zero, this means that there 
is an open string operator $\psi$ satisfying
\begin{equation}
\eqlabel{cancelwarner}
\phi^{(1)}|_{\del\Sigma} = \{G^+_{\rm bdry},\psi^{(1)}\}
\end{equation}
where $G^+_{\rm bdry}$ is the boundary part of the supercharge. Hence
by adding the boundary term
\begin{equation}
\eqlabel{bodef}
\delta S_{\rm bdry} = -\int_{\del\Sigma} \psi^{(1)}
\end{equation}
to the action, we can cancel the Warner term in the susy variation,
and the brane deforms with the closed string background.

On the other hand, if $m_0(\phi)\neq 0\in {\rm Ext}^2(B,B)$, we will
not be able to deform the brane linearly with the background. However,
note that by Serre duality (or worldsheet CPT), ${\rm Ext}^2(B,B)
\cong {\rm Ext}^1(B,B)$, and it can happen that $m_0(\phi)$ is
in the image of a higher product $m_n$ for some $n$. In this case, 
by locking together an obstructed boundary deformation with the bulk 
deformation, we can still deform the brane with the closed string.

The basic example to keep in mind is when there is just one bulk 
deformation, $t$, and one boundary deformation, $u$. The superpotential 
is now a function of $u$ and $t$, where we can treat the latter as 
a parameter.
\begin{equation}
\eqlabel{generically}
\calw=\calw(u;t) = \mu_0 tu - \frac  {\mu_n} {n+1} {u^{n+1}} + \calo(u^{n+2})
\end{equation}
where $\mu_0$ and $\mu_n$ are constants. If $\mu_0,\mu_n\neq 0$, then for 
$t=0$, $u$ is obstructed at order $n$, while for small $t\neq 0$, there 
are $n$ vacua $u\sim t^{1/n}$, and $u$ is massive around each of them.

Let us bring this discussion to the point. If the D-brane has no marginal 
deformations, then the bulk-to-boundary obstruction map must be zero 
because ${\rm Ext}^2=0$. If there is a marginal deformation, and a 
non-trivial $m_0$, but the D-brane does not obstruct the bulk deformation, 
then $\mu_n$ must be non-zero for some smallest $n\ge 2$, and the marginal 
boundary direction is lifted by a small bulk deformation.

\subsection{Closed string holomorphic anomaly}
\label{cshan}

As we have mentioned, topologically twisted $\caln=2$ SCFTs can be coupled to 
2d (topological) gravity by identifying the supercharges and their conjugates
with BRST operators and antighosts of a critical bosonic string in which
the ghost and matter fields do not decouple. Indeed, the algebra \eqref{algebra} 
is all that is needed to define string amplitudes by integration over the moduli 
space of Riemann surfaces.

If $\calmc g$ denotes the moduli space of Riemann surfaces of genus $g\ge 2$, 
and $\mu_a$, $a=1,\ldots,3g-3$ the Beltrami differentials, we define the 
topological string amplitude at genus $g$ by the formula,
\begin{equation}
\eqlabel{Fgclosed}
\Fc g = \int_{\calmc g}  [dm] \; \bigl\langle \prod_{a=1}^{3g-3} 
\bigl(\int \mu_a G^-\bigr)
\bigl(\int \mu_\ab \Gb^-\bigr) \bigr\rangle_{\Sigma_g}
\end{equation}
where $\mu_a G^-\equiv (\mu_a)^z_{\;\zb} G^-_{zz}$ denotes the contraction of the 
Beltrami's with the antighosts, and $\langle\cdots\rangle_{\Sigma_g}$ the 2d 
field theory correlator on the worldsheet $\Sigma_g$. By this definition, 
the $\Fc g$ become sections of the line bundle $\call^{2g-2}$ over the CFT 
moduli space $M$. The definition has to be modified slightly for $g=0$ and
$g=1$ because of the presence of ghost zero modes as usual. 

\begin{figure}[t]
\psfrag{kbar}{$\int \phi_\ib$}
\epsfig{height=5cm,file=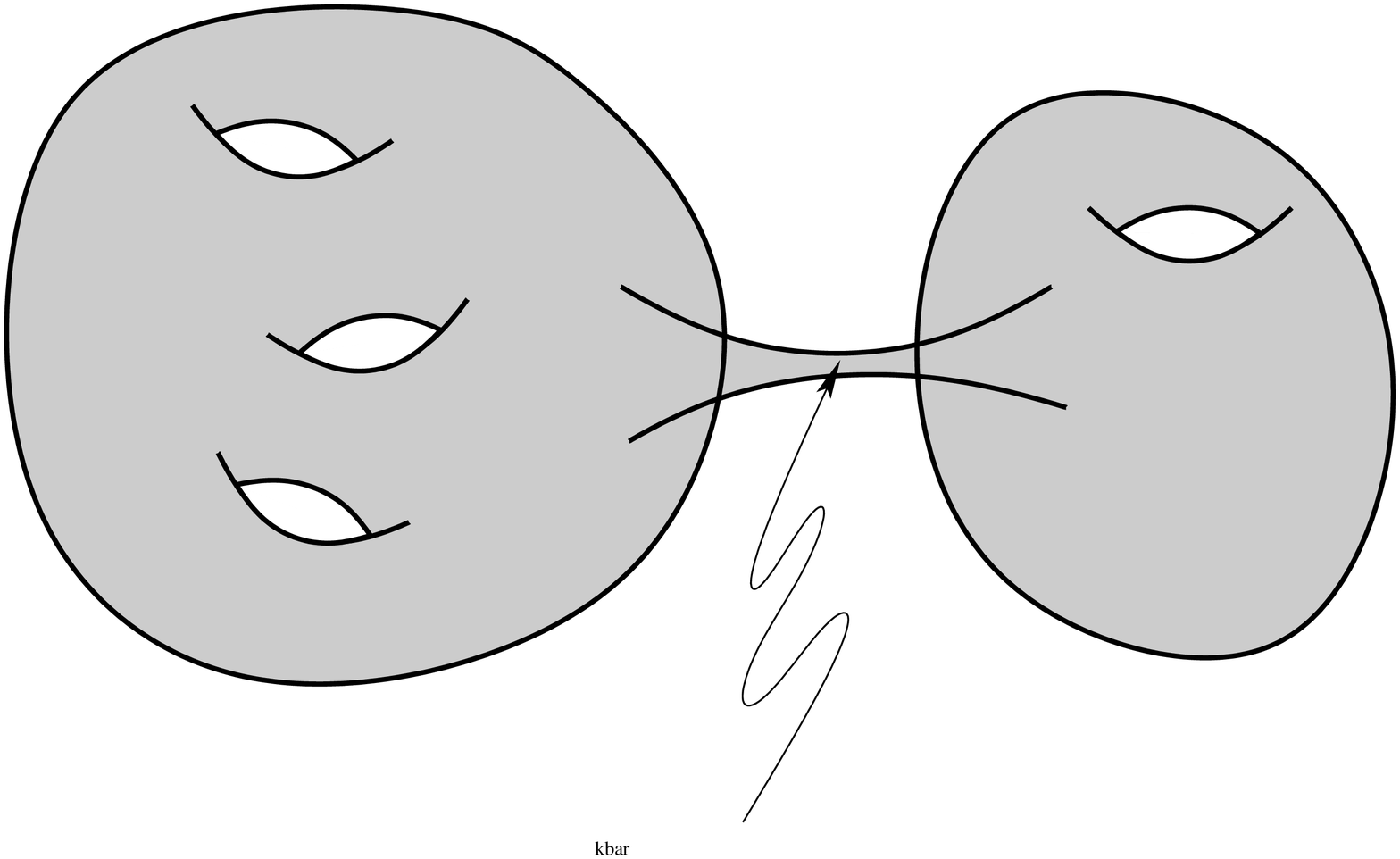}
\qquad
\epsfig{height=5cm,file=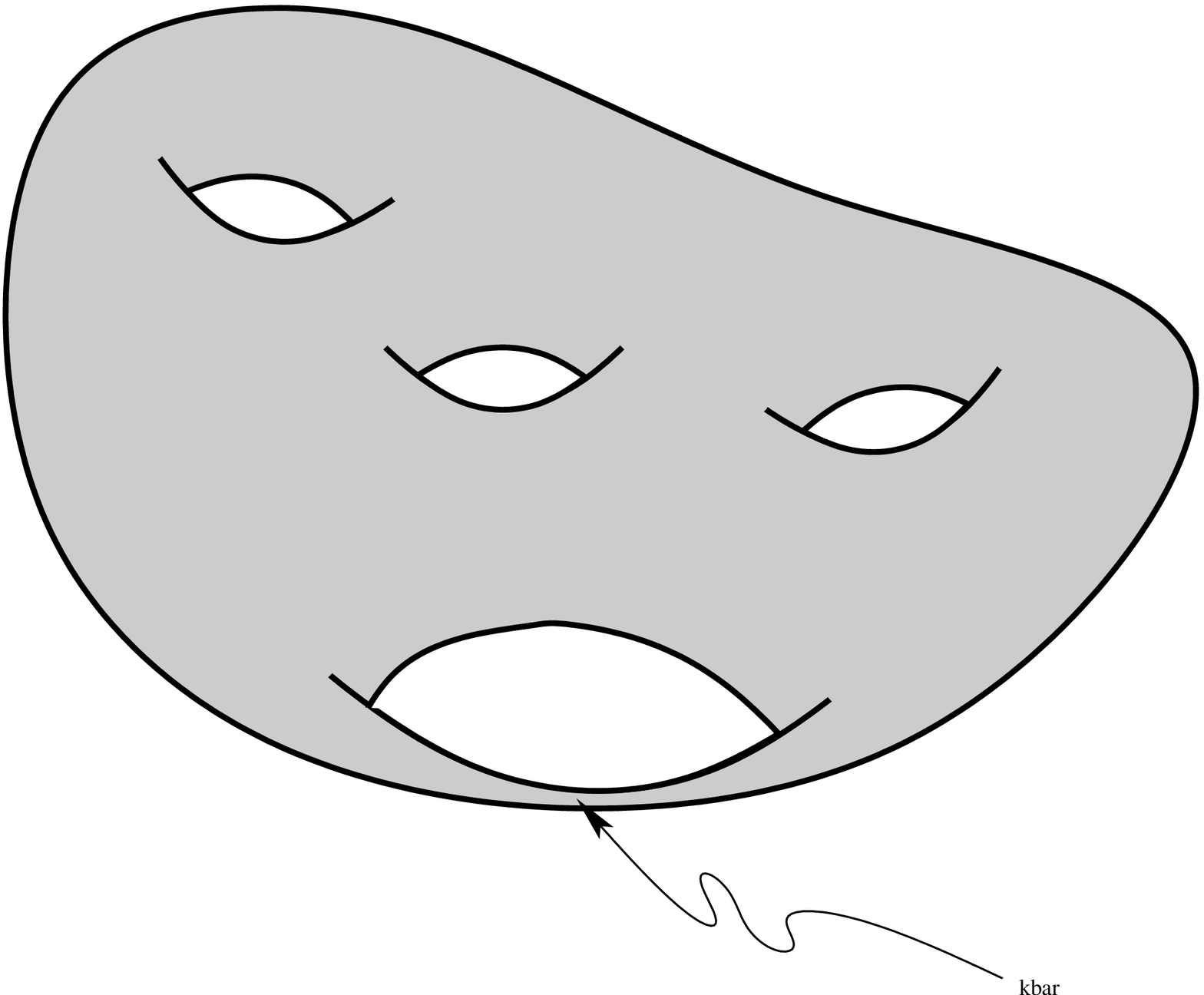}
\caption{The two degenerations that contribute to the RHS of the
holomorphic anomaly for closed strings.}
\label{degenbcov}
\end{figure}

To derive the holomorphic anomaly equation, BCOV consider an infinitesimal 
deformation of the action by
\begin{equation}
t^i \int \phi_i^{(2)} + t^\ib \int\phi_\ib^{(2)}
\end{equation}
and differentiate \eqref{Fgclosed} with respect to $t^\ib$. This leads to the
insertion of $\int\phi_\ib^{(2)}=\int d^2z \sqrt{h} \{G^+,[\Gb^+,\phi_\ib]\}$ 
in the topological correlator. By contour deformation, one can 
move the action of the BRST operators to the Beltrami differentials folded 
with the anti-ghosts. By using the zero-mode algebra \eqref{algebra}, the 
contractions are converted into differentials $\del\langle\cdots\rangle$, 
$\delbar\langle\cdots\rangle$ on $\calmc g$,
\begin{equation}
\frac{\del}{\del t^\ib} \Fc g =\int_{\calmc g} [dm]
\sum_{a,\ab=1}^{3g-3} 4 \frac{\del^2}{\del m^a\del m^\ab} \bigl\langle \int \phi_\ib 
\prod_{\topa{a'\neq a}{\ab'\neq\ab}}
\bigl(\int\mu_{a'} G^-\bigr)
\bigl(\int\mu_{\ab'} \Gb^-\bigr) \bigr\rangle_{\Sigma_g}
\end{equation}
Thus the anti-holomorphic derivative of $\Fc g$ has been converted into
the integral of a total derivative over $\calmc g$. Naively, this is
zero, but a careful analysis reveals that the correlator 
$\langle\int\phi_\ib \cdots\rangle_{\Sigma_g}$ exhibits singularities 
around certain ``boundary components'' of $\calmc g$, which correspond 
to our Riemann surface degenerating in various ways. Two types of 
degenerations are relevant for the closed string. The Riemann
surface can split in two components by developing a long tube, or
a handle can pinch without leading to a disconnected $\Sigma_g$. See 
fig.\ \ref{degenbcov}.

In the derivation, one has to pay close attention to the location of
the anti-chiral field in the limit where $\Sigma_g$ degenerates. As 
it turns out, the whole contribution comes from the region where 
$\phi_\ib$ is sitting on the long tube. Thus, one is more nearly 
considering a Riemann surface which pinches at {\it both ends} 
of the tube. In the limit, the pinches are each repaired by inserting
complete sets of chiral fields, $\phi_j$, $\phi_k$ on the lower-genus
Riemann surface. The long tube is replaced with the three-point function 
on the sphere with insertion of anti-chiral fields, $\phi_\ib$, $\phi_\jb$, 
$\phi_\kb$. This reduces to the anti-holomorphic Yukawa coupling 
$C_{\ib\jb\kb}$. The connection between the various pieces of $\Sigma_g$
occurs via the inverse of the topological metric $g^{\jb j}$, $g^{\kb k}$.

Another important aspect of the derivation is that (for $g\ge 2$!), the
sums at the location of the pinches are actually {\it only over the
marginal fields}, \ie, those of charge $(q,\qb)=(1,1)$. This restriction
arises from combining $U(1)$ charge conservation on the sphere, $q_i+q_j+
q_k=3$, with the fact that $\phi_a^{(2)}=0$ for $q_a=0$. Namely,
integrated insertion of the identity operator leads to a vanishing
contribution. Since $q_i=1$ already, the only remaining possibility is 
$q_j=q_k=1$.

Taken together, one obtains the holomorphic anomaly equation for $\Fc g$ 
($g\ge 2$) as derived in BCOV 
\begin{equation}
\eqlabel{bcov}
\del_\ib \Fc g = \frac 12 \sum_{g_1+g_2=g} C_{\ib}^{jk}
\Fc {g_1}_j \Fc {g_2}_k + \frac 12
C_{\ib}^{jk} \Fc {g-1}_{jk} \,,
\end{equation}
where the $\frac 12$ is a symmetry factor, and $C_\ib^{jk}\equiv
C_{\ib\jb\kb} g^{\jb j} g^{\kb k}= C_{\ib\jb\kb}\ee^{2K} G^{\jb j} 
G^{\kb k}$. The $\Fc g$ with subscripts are the topological string 
amplitudes with insertion of the corresponding chiral fields, and
are defined by
\begin{equation}
\Fc g_{i_1,\ldots, i_n} = \int_{\calmc g} [dm]\;
\bigl\langle
\int \phi_{i_1}^{(2)} \cdots \int \phi_{i_n}^{(2)}
\prod_{a=1}^{3g-3} 
\bigl(\int \mu_a G^-\bigr)
\bigl(\int \mu_\ab \Gb^-\bigr)
\bigr\rangle
\end{equation} 
As also shown in BCOV, the amplitudes with insertions can be obtained from 
the partition functions, $\Fc g$, by covariant differentiation,
\begin{equation}
\Fc g_{i_1,\ldots,i_{n+1}} = D_{i_{n+1}} \Fc g_{i_1,\ldots,i_n}
\end{equation}
where again $D$ is the Zamolodchikov-K\"ahler covariant derivative on 
${\rm Sym}^n T^*M\otimes\call^{2g-2}$.

Note that in \eqref{bcov}, the sum is restricted to $g_i\ge 1$, which
is a consequence of the vanishing of the sphere one and two-point function.
Namely, we only encounter {\it stable degenerations}, with $3g_i-3+n_i>0$
for $i=1,2$, where $n_i\equiv 1$ is the number of marked points on the $i$-th
component.

\subsection{More on the vacuum bundle. D-branes as normal functions}
\label{more}

\begin{figure}[ht]
\begin{center}
\psfrag{state}{$|0\rangle_{\rm open}$}
\epsfig{height=3cm,file=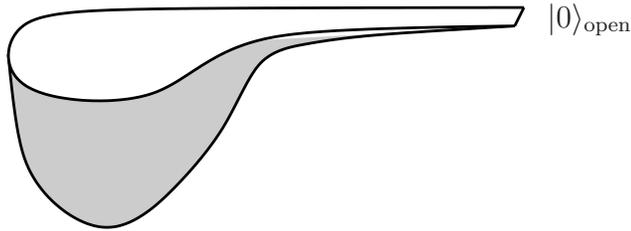}
\caption{The path-integral on the disk with a long strip attached 
defines a canonical open string vacuum.}
\label{attached}
\end{center}
\end{figure}

Let us begin this subsection with a minor comment, which will gain 
some importance in the one-loop holomorphic anomaly in subsection 
\ref{oneloophan}. As we have recalled above, topological string 
amplitudes at genus $g$ are sections of the non-trivial line bundle 
$\call^{2g-2}$ over moduli space $M$. This arises because the 
ambiguity in normalizing the twisted path-integral on the worldsheet 
is determined by the choice of a canonical closed string vacuum 
$e_0=|0\rangle_{\rm closed}$. It is natural to ask whether the 
presence of boundaries could introduce new ambiguities, and lead to 
additional twisting. In fact, there are no new ambiguities, as can be 
seen from considering the topological path-integral on a disk with a 
long strip attached (Fig.\ \ref{attached}). Given $|0\rangle_{\rm closed}$,
this defines a canonical open string vacuum $|0\rangle_{\rm open}$, with 
a canonical normalization. As a consequence, open topological string 
amplitudes at genus $g$ with $h$ boundaries will be sections of 
$\call^{2g+h-2}$ over $M$.

To continue the discussion, it will be helpful to have in mind a more 
concrete geometric realization of the topological string. So let us 
consider the B-model on a family of Calabi-Yau manifolds, all denoted
by $Y$. The ground states are identified with the cohomology of $Y$ via
\begin{equation}
\calh^{p,q} \cong H^{3-p,q}(Y)
\end{equation}
while the $(c,c)$-ring structure is determined from the identification
\begin{equation}
H^{3-p,q}(Y) \cong H^q(\Lambda^p TY)
\end{equation}
given by contraction with the holomorphic 3-form, $\Omega$. The moduli
space $M$ is the space of complex structure deformations of $Y$. As $Y$
varies over $M$, the middle dimensional cohomology groups $H^3(Y)$
fit together into a holomorphic vector bundle, which is precisely the
vacuum bundle $\calv$ we have discussed in subsection \ref{vacgeo1} 
above. The decomposition \eqref{decompose} is now
\begin{equation}
\eqlabel{central}
\calv_m = H^{3,0}(Y) \oplus H^{2,1}(Y)\oplus H^{1,2}(Y) \oplus H^{0,3}(Y)
\end{equation}
An important point is that the decomposition \eqref{central} is not compatible
with the holomorphic structure on $\calv$. Instead, consider the {\it Hodge 
filtration} on $H^3(Y)$,
\begin{equation}
\eqlabel{filtration}
H^{3,0}(Y)=F^3H^3(Y)\subset F^2H^3(Y)\subset F^1H^3(Y)\subset F^0 H^3(Y)=H^3(Y)
\end{equation}
where
\begin{equation}
F^qH^3(Y) = \oplus_{q'\ge q} H^{q',3-q'}(Y) 
\end{equation}
is the space of three-forms with at least $q$ holomorphic indices. The
$F^qH^3(Y)$ do fit together into holomorphic subbundles of $\calv$ over
$M$. In particular, $F^3H^3(Y)=H^{3,0}(Y)$ is identified with our
canonical line bundle $\call$.

The topological metric on $\calv$ is up to a sign, $(-1)^q$, the symplectic 
pairing between $H^{3-q,q}(Y)$ and $H^{q,3-q}(Y)$, while the Zamolodchikov 
metric is identified with the Weil-Petersson metric on $M$
\begin{equation}
\eqlabel{g00}
G_{i\jb} = \del_i\del_\jb K\,,\qquad\text{where } \;
K = -\log \ii\int_Y\Omegab\wedge\Omega
\end{equation}
The structure constants of the chiral ring are given by
\begin{equation}
\eqlabel{structure}
C_{ijk} = - \int_Y \Omega\wedge\del_i\del_j\del_k\Omega
\end{equation}
The vacuum bundle $\calv$ comes equipped with a real structure, which
is induced from the embedding $H^3(Y;\reals)\subset H^3(Y;\complex)$.
Complex conjugation acts by exchanging $H^{q,3-q}$ with $H^{3-q,q}$
and corresponds on the worldsheet to the CPT operator $\Theta$. 

It can hardly be overemphasized that the starting point for much of
special geometry and in fact the entire story of the holomorphic anomaly 
is the competition between the holomorphicity of the filtration 
\eqref{filtration}, which would make topological string amplitudes
holomorphic, and the reality of the decomposition \eqref{central},
which is preferred for maintaining unitarity of the worldsheet CFT.
We will see that this is crucial when we add boundaries as well.

Not only do we have a real structure on $H^3(Y;\complex)\supset
H^3(Y;\reals)$, but we also have an {\it integral structure} from
the embedding $H^3(Y;\zet)\subset H^3(Y;\reals)$. The Gauss-Manin 
connection on $H^3(Y;\complex)$ can be characterized by the fact that 
it preserves this integral structure. Namely, any section of $H^3(Y;\zet)$ 
is flat with respect to the Gauss-Manin connection. On the Hodge 
filtration, the Gauss-Manin connection satisfies Griffiths 
transversality
\begin{equation}
\eqlabel{transverse1}
\nabla F^qH^3(Y)\subset F^{q-1} H^3(Y)
\end{equation}
If $\Gamma\in H_3(Y;\zet)$ is an integral 3-cycle, it defines 
an element in $H^3(Y;\zet)$ by integration against
3-forms and duality. Of importance are the periods of the
holomorphic three-form
\begin{equation}
\eqlabel{period1}
\Pi(z) = \langle\phi_0,\Gamma\rangle = \int_\Gamma\Omega(z)
\end{equation}
where $z=(z^i)_{i=1,\ldots, n}$ is some collection of local 
coordinates on $M$. The pairings with the $(2,1)$-forms $\chi_i$
can be obtained by differentiation,
\begin{equation}
\eqlabel{period2}
D_i\Pi(z) = \del_i\Pi(z)+\del_i K \Pi(z) 
= \langle\phi_i,\Gamma\rangle = \int_\Gamma \chi_i^{(2,1)}(z)
\end{equation}
while the overlaps with the $(1,2)$ and $(0,3)$-forms follow (for
instance) by complex conjugation.

It will be useful for us to introduce at this stage the so-called
{\it Griffiths intermediate Jacobian}. At any point in moduli space, 
we consider the complex torus
\begin{equation}
\eqlabel{intermediate}
J^3(Y) = H^{1,2}(Y)\oplus H^{0,3}(Y)/H^3(Y;\zet)
= H^3(Y;\complex)/(F^2H^3(Y) + H^3(Y;\zet))
\end{equation}
(The underlying real torus is $H^3(Y;\reals)/H^3(Y;\zet)$, the
complex structure is determined by the complex structure on $Y$.)
Because of the holomorphicity of the filtration \eqref{filtration},
the $J^3(Y)$ fit together into a holomorphic family of complex
tori, known as the intermediate Jacobian fibration.

We can now begin to ask more precisely how D-branes fit into 
the framework of special geometry and the vacuum bundle over $M$.
A-branes first.

From the worldsheet perspective, A-branes are essential to define 
the integral structure on the vacuum bundle. Indeed, while the 
worldsheet CPT operator defines the real structure, it does 
not allow the selection of an integral lattice inside of it. 
But consider two A-branes $A$ and $A'$. By general principles, 
we can express the Witten index in the Hilbert space of 
$A$--$A'$-strings via the overlaps of the boundary states with 
the Ramond ground states, which for A-branes must be taken from 
the B-model as we have explained in subsection \ref{first}
\begin{equation}
\eqlabel{witten}
{\rm tr}_{A-A'}(-1)^F = \langle A|e_a\rangle\eta^{ab}\langle e_b|A'\rangle
\end{equation}
where $e_a$ is some basis of $\calv$, and $\eta^{ab}$ the inverse 
topological metric. Geometrically, A-branes wrap (special) Lagrangian 
three-cycles in $Y$. Their class in $H_3(Y;\zet)$ defines the overlap
with $\calv$ via \eqref{period1}, \eqref{period2}, and their 
(integral!) intersection number computes the Witten index 
\eqref{witten}. Thus, properly normalized A-brane boundary states
define integral sections of $\calv$.

B-branes are more subtle. (Useful references for the geometric
statements that will follow below include \cite{cime,voisinbook}.) 
As we have mentioned before, the modern perspective is 
that D-branes are mathematically well accommodated in certain categories 
endowed with extra structure, such as the $A_\infty$-structure we have 
discussed in subsection \ref{first}. For our B-model on $Y$, the 
relevant category is the bounded derived category of coherent sheaves, 
$D^b(Y)$. Essentially, this includes D-branes wrapped on 
even-dimensional, holomorphic cycles carrying holomorphic vector
bundles, as well as all possible ``topological bound states'' of those
that can be obtained by ``topological tachyon condensation''. 

Our main goal is to extract holomorphic information from objects in 
the B-brane category, and to relate it to the vacuum bundle $\calv$. 
It is in any case clear already that the {\it topological} 
classification of B-branes involves ground states 
from the other model, which are not contained in $\calv$.  For 
reasons that will become completely clear only in subsection 
\ref{openhan}, we want to restrict ourselves to B-branes whose 
overlaps with those ground states with $q_{\it rgs}=
-\qb_{\it rgs}$ vanish. The point is that if those overlaps don't 
vanish, one is in danger that the corresponding A-model deformations, 
which are BRST exact in the B-model, will not decouple in loop 
computations. For this reason, and because it seems plausible that 
analogous effects can be achieved by (topological) orientifolds, we 
refer to this condition as ``tadpole cancellation''. 

Physically, all the holomorphic information one expects to extract 
from the open string at tree-level is captured in the spacetime 
superpotential $\calw$ for the massless fields on the brane. The
interpretation of the superpotential in $A_\infty$-categories is 
quite well understood, but depends to a large extent on gauge-fixing
data of the open string field theory \cite{wittencs,calin}. 
Gauge invariant physical information contained in the superpotential
is for example the tension of BPS domainwalls, $\calt=\Delta\calw$.
Since such domainwalls carry no topological D-brane charge, they
are precisely the physical objects satisfying the tadpole
cancellation condition from the previous paragraph. We will henceforth
restrict ourselves to such configurations.

Consider for instance wrapping of a D5-brane on a holomorphic curve 
$C\subset Y$. This B-brane carries no topological charge if its class 
in $H_2(Y;\zet)$ vanishes. It can then arise as a domainwall between 
two D5-branes wrapped on two different holomorphic curves $C_+$ and 
$C_-$ in the same class if $C=C_+-C_-$ holomorphically. The tension
of the BPS domainwall is \cite{wittenqcd}
\begin{equation}
\eqlabel{tension}
\calt = \int_{\Gamma} \Omega
\end{equation}
where $\Omega(z)$ is the holomorphic three-form and $\Gamma$ is a three-chain
in $Y$ with boundary $\del\Gamma=C=C_+-C_-$. The domainwall tension
depends on complex structure moduli both explicitly through the holomorphic 
three-form, as well as implicitly through the position of the curve $C$,
which must vary in order to remain holomorphic as we vary the complex 
structure. At this stage, we also allow dependence of $\calt$ on any
moduli of $C$ for fixed complex structure of $Y$, but we will drop this 
freedom in the next subsection.

Because the holomorphic three-form is unique up to scale, formula \eqref{tension} 
is well-defined even if we think of $\Omega$ just as a representative of 
a cohomology class in $H^{3,0}(Y)$. Because $\Gamma$ has a boundary, we could 
not integrate an arbitrary cohomology class over it. However, next
to $\Omega$, we can do one more. Consider a class $[\chi^{(2,1)}]$ in
$H^{2,1}(Y)$. It is an elementary fact from Hodge theory that
$F^pH^k\cong(F^pA^k)^c/dF^pA^{k-1}$, where $F^pA^k$ are the $k$-forms on
$Y$ with at least $p$ holomorphic indices, $(\cdot)^c$ refers to closed
forms, and $d$ is the total differential. Thus we can represent $[\chi^{2,1}]$ 
by a closed form $\chi^{(2,1)}\in F^2 A^3$ and define the integral
\begin{equation}
\eqlabel{integral}
\int_{\Gamma} \chi^{(2,1)}
\end{equation}
The integral does not depend on the choice of representative since under 
$\chi^{(2,1)}\to\chi^{(2,1)}+ d\xi^{(2,0)}$, where $\xi^{(2,0)}$ is a 
$(2,0)$-form, the integral changes by $\int_{\del\Gamma} \xi^{(2,0)}$, 
and since $\del\Gamma=C$ is a holomorphic curve, the integral of a 
$(2,0)$-form over it vanishes.

The properties we have just described are part of the definition of a 
{\it Poincar\'e normal function}, in the sense of Griffiths \cite{griffiths1} 
(see \cite{cime} for a pedagogical introduction). Formally, a normal
function $\nu$ is a holomorphic section of the intermediate
Jacobian \eqref{intermediate} satisfying the infinitesimal condition 
for normal functions, or Griffiths transversality, defined as follows.
If $\nu$ is any holomorphic section of $J^3(Y)=H^3(Y)/(F^2H^3+H^3(\zet))$, 
one can choose a lift of $\nu$ as a holomorphic section $\tilde\nu$ of 
$H^3(Y)$. Then we can apply the Gauss-Manin connection $\nabla$ to 
$\tilde\nu$, and Griffiths transversality for normal functions is the 
statement
\begin{equation}
\eqlabel{normtrans}
\nabla\tilde\nu \in F^1H^3 
\end{equation}
Instead of showing that this condition is independent of the lift
(which it is), let us verify that a family of holomorphic curves indeed 
defines a normal function. Note that by duality, we can identify 
$H^3/F^2H^3$ with $(F^2H^3)^*$ and $J^3(Y)$ with $(F^2H^3)^*/H_3(Y;
\zet)$. Correspondingly, the integrals \eqref{tension}, \eqref{integral} 
define an element of $(F^2H^3)^*=(H^{3,0}\oplus H^{2,1})^*$, and
the three-chain $\Gamma$ with $\del\Gamma=C$ is only defined up to a 
three-cycle in $H_3(Y;\zet)$. Thus, $C$ defines a section of the 
intermediate Jacobian. Finally, Griffiths transversality \eqref{normtrans} 
follows from the observation that when we vary the complex structure
of $Y$, we can describe the first order variation of $C$ by a normal 
vector $n\in N_{C/Y}$. If $\delta\Gamma$ is the corresponding first
order variation of $\Gamma$, one has
\begin{equation}
\eqlabel{equivalent}
\int_{\delta\Gamma} \Omega = \int_C \Omega(n) = 0
\end{equation}
which again vanishes by type considerations since $C$ is holomorphic.
This is equivalent to $\langle\Omega,\nabla\tilde\nu\rangle=0$, and 
hence to \eqref{normtrans}.

Normal functions also make sense for holomorphic vector bundles, and 
by splitting distinguished triangles can be defined for the entire 
derived category $D^b(Y)$. The essential device that makes this possible 
is the notion of algebraic or holomorphic second Chern class. Given 
for example a holomorphic vector bundle, we can equip it with a 
hermitian metric, and thus specify a connection, $A$, whose curvature 
$F=dA+A\wedge A$ is of type $(1,1)$. The second Chern form is
$c_2(A) = {\rm tr} F\wedge F$ and defines a cohomology class in 
$H^4(Y;\zet)$. If $[c_2(A)]=0$, one may write $c_2(A)=d {\it CS}(A)$, where 
${\it CS}(A)$ is the Chern-Simons form. This way, we identify the domainwall 
tension with Witten's holomorphic Chern-Simons functional \cite{newissue,
wittencs,doth},
\begin{equation}
\eqlabel{holcs}
\calt = \int {\it CS}(A)\wedge \Omega
\end{equation}
and indeed one can show that it depends only on the holomorphic class
of the vector bundle.

Further details on the relation of D-branes to normal functions, 
with an important example, appear in \cite{mowa}.

\subsection{Infinitesimal invariant and holomorphic anomaly on the disk}
\label{infinvhan}

We have just seen that topologically trivial B-branes are holomorphically
captured by a normal function, namely a holomorphic section of the intermediate
Jacobian \eqref{intermediate} satisfying Griffiths transversality 
\eqref{normtrans}. This association is known as the {\it Abel-Jacobi map}. 
It is worthwhile pointing out that in general, one can also consider 
intermediate Jacobians for 0-cycles, $J^1(Y)$, and for four-cycles,
$J^5(Y)$. However, if $Y$ is simply connected (which we assume), 
those Jacobians, known as the Albanese and Picard variety, respectively, 
vanish. The Abel-Jacobi map was first used used for open string disk 
instanton computations (on non-compact Calabi-Yau) by Aganagic and 
Vafa \cite{agva}. Early speculations on the relevance of the Abel-Jacobi 
map to mirror symmetry appear in \cite{doma}. In this subsection, we 
study the relation between the Abel-Jacobi map for B-branes and the 
vacuum bundle $\calv$, especially at the infinitesimal level. 

It is clear from the previous subsection that a normal function in 
itself cannot completely describe the topological boundary state of 
a B-brane. Granting a lift of the $H^3(Y;\zet)$ ambiguity, $\nu$ 
only defines the $(0,3)$ and $(1,2)$ components of an element of 
$H^3(Y)$, and this only in the quotient. To get an actual state in 
$\calv$, we need a lift $\tilde\nu$.

A little thought reveals that there is in fact a very natural lift
of $\nu$ to all of $\calv$, dictated by worldsheet CPT invariance.
Since the latter is simply complex conjugation acting on $H^3(Y)$, we 
see that at the level of the pairing $\langle\Omega,\nu\rangle=
\int_\Gamma\Omega$, \eqref{tension}, we are defining the lift by
\begin{equation}
\eqlabel{reallift}
\langle\Omegab,\tilde\nu\rangle = \int_\Gamma\Omegab 
= \overline{\int_\Gamma\Omega} = \overline{\langle\Omega,\tilde\nu\rangle}
\end{equation}
and similarly for the $(1,2)$-forms. We will henceforth denote this
real lift of the normal function also by $\nu$.

Before studying the full consequences of this identification, let us
finally clarify our intent to neglect open string moduli that has 
been lingering since (F1) in the introduction. We have seen already 
in subsection \ref{first} that if bulk deformations are unobstructed
by the D-brane and the obstruction map $m_0$ is non-zero, we can
remove open string moduli by a small bulk deformation. 

To deal with the assumption that $m_0$ is non-trivial, consider a 
family of homologically trivial B-branes $B(w)$, which as a function 
of some local parameter $w$ are all holomorphic with a fixed complex 
structure of $Y$. We can define the Abel-Jacobi map ${\it AJ}(w)\in 
J^3(Y)$. By considerations similar to those around \eqref{equivalent}, 
one can show that the first order variation of ${\it AJ}(w)$ satisfies
\begin{equation}
d_w {\it AJ}(w) \in F^1H^3(Y)/F^2H^3(Y)\cong H^{1,2}(Y)
\end{equation}
This is similar to \eqref{normtrans}, except that we now vary only
the brane for fixed complex structure of $Y$. Since the tangent space
to moduli of $B(w)$ is ${\rm Ext}^1(B,B)$ as reviewed in subsection
\ref{first}, the infinitesimal Abel-Jacobi map is more abstractly a
map
\begin{equation}
\eqlabel{infaj}
\alpha: {\rm Ext}^1(B,B)=\calh^1_{\rm open} \to H^{1,2}(Y)= \calh^{2,2}_{\rm closed}
\end{equation}
Diagrammatically, by using the closed string topological metric, we can
identify $\alpha$ with the two point function on the disk with one
boundary and one bulk insertion, see Fig.\ \ref{obstruct}. Referring
back to subsection \ref{first}, we see that {\it the infinitesimal 
Abel-Jacobi map is nothing but the dual of the bulk-to-boundary 
obstruction map \eqref{obmap}}, $\alpha=m_0^*$. Thus, if $m_0$ 
vanishes, the image of $B(w)$ in the intermediate Jacobian is 
independent of $w$, and, if the brane does not obstruct the bulk, 
the corresponding normal function will also not depend on $w$.

When the B-brane is a holomorphic vector bundle, these statements
are reflected in the fact that the holomorphic Chern-Simons functional
\eqref{holcs} is constant on unobstructed families of holomorphic 
connections \cite{doth}. In fact, the holomorphic Chern-Simons functional
(or, more generally, the open string field theory) and its quantization
encodes the entire deformation and obstruction theory for B-branes 
\cite{clemens}. We are here only concerned with its most elementary 
application.

An extremely useful concept attached to normal functions in the
context of infinitesimal variation of Hodge structure is the 
so-called Griffiths' infinitesimal invariant. It was first
considered by Griffiths in \cite{griffiths2}, and later refined
by Voisin \cite{voisin} and Green \cite{green}. If one insists on
holomorphicity, defining the infinitesimal invariant requires some 
ingenuity, because one cannot quite do it without choosing a lift 
of $\nu$ to $H^3(Y)$ (see \cite{cime}). But since we have given 
up on holomorphicity long ago, and work with the real physical 
lift \eqref{reallift}, we can be more pedestrian. Our main goal
is to explain the identification of the infinitesimal invariant
with the disk two-point function, see (F3) in the introduction.

Consider a real normal function $\nu$, and expand in a basis of
the vacuum bundle, see eq.\ \eqref{newbasis},
\begin{equation}
\nu = \nu^0 e_0+\nu^i e_i + \nu^\ib e_\ib + \nu^\0b e_\0b
\end{equation}
where reality means $\nu^\ib=\overline{\nu^i}$, $\nu^\0b=
\overline{\nu^0}$. The domainwall tension is of course 
$\calt=\langle\Omega,\nu\rangle=\langle e_0,\nu\rangle=\nu^\0b 
g_{0\0b}$. By utilizing the explicit form of the connection matrices 
in subsection \ref{vacgeo1}, we find
\begin{multline}
\eqlabel{explderiv}
\nabla_i\nu=(\del_i\nu^0-\del_i K\nu^0)e_0 +
\bigl(\del_i\nu^l + g^{\kb l}\del_i g_{m\kb} \nu^m -\delta_i^l\nu^0\bigr)
e_l + \\ \bigl(\del_i\nu^\lb - C_{im}^{\;\;\;\;\lb} \nu^m\bigr) e_\lb
+\bigl(\del_i\nu^\0b - G_{i\mb}\nu^\mb\bigr) e_\0b
\end{multline}
Griffiths transversality for normal functions is the
statement $\langle\Omega,\nabla\nu \rangle=0$, which translates into
\begin{equation}
D_i\calt =\del_i\calt+\del_i K \calt =  \nu_i = g_{i\jb} \nu^\jb
\end{equation}
The Griffiths' infinitesimal invariant can now be defined as
the following tensor in $(T^*M)^2 \otimes \call^{-1}$,
\begin{equation}
\Delta_{ij} := -\langle\nabla_i \Omega , \nabla_j \nu\rangle
\end{equation}
By using Griffiths transversality and the compatibility of the
Gauss-Manin connection with the symplectic metric, this
is equivalent to 
\begin{equation}
\Delta_{ij} = \langle \Omega , \nabla_i\nabla_j\nu\rangle
\end{equation}
which makes it obvious that $\Delta_{ij}$ is symmetric in $i$
and $j$, \ie, $\Delta_{ij}\in {\rm Sym}^2(T^*M)\otimes\call^{-1}$.
From \eqref{explderiv}, we find explicitly
\begin{equation}
\eqlabel{infinv}
\Delta_{ij} = D_iD_j\calt - C_{ijk} g^{\kb k} D_\kb \overline\calt
\end{equation}
From the definition, it is clear that $\Delta_{ij}$ vanishes 
identically if $\calt$ is a period of the holomorphic three-form 
over a (closed) three-cycle. To verify this, one has to be careful to
insert an actual real period, and not just an arbitrary complex 
solution of the Picard-Fuchs equation. These notions are not 
equivalent since \eqref{infinv} is not holomorphic in $\calt$.
But in any event, we see that the infinitesimal invariant does
not depend on how we choose to lift the $H^3(Y;\zet)$ ambiguity
in the definition of the normal function. It is also invariant
under monodromies in the complex structure moduli space, for
the same reason.

To show the identification of $\Delta_{ij}$ with the disk two-point 
function, we will make use of the holomorphic anomaly. From 
\eqref{infinv} and the special geometry relations, it is not hard 
to see that our infinitesimal invariant satisfies the distinctive 
equation
\begin{equation}
\eqlabel{diskhan}
\del_\ib \Delta_{jk} =-C_{jkl} g^{\lb l}D_\ib D_\lb\overline{\calt}
+ C_{jkl} C_{\ib\lb\mb} g^{\lb l} g^{\mb m} D_m\calt 
= - C_{jkl} g^{\lb l} \Delta_{\ib\lb}
\end{equation}
where $\Delta_{\kb\lb}=\overline{\Delta_{kl}}$.
\begin{figure}[ht]
\begin{center}
\psfrag{phii}{$\phi_j(0)$}
\psfrag{phij}{$\int \phi_k^{[1]}(r)$}
\epsfig{height=3.5cm,file=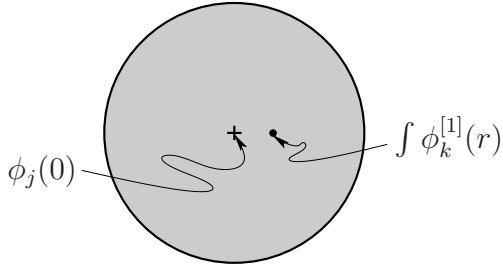}
\caption{The disk amplitude with two bulk insertions.}
\label{twopoint}
\end{center}
\end{figure}
On the other hand, the topological string amplitude on the disk
with two bulk insertions is defined by
\begin{equation}
\tilde \Delta_{jk} = \int_0^1 dr \langle \phi_j(0) \phi_k^{[1]}(r)\rangle_{(0,1)}
\end{equation}
where one of the insertions is fixed at $0$, and we are integrating
over the radial position of the one-form descendant $\phi_k^{[1]}
=\frac 12[G^--\Gb^-,\phi_k]$ of the other. Taking a derivative of $\tilde\Delta_{jk}$ 
in the anti-holomorphic direction brings down the anti-chiral insertion
$\int \phi_\ib^{(2)}= \int d^2z\sqrt{h}\frac 12\{G^++\Gb^+,[G^+-\Gb^+,
\phi_\ib]\}$. This is BRST exact in the presence of the boundary, and 
similarly to the derivation in subsection \ref{cshan}, we can move
the BRST operator to the chiral insertion, where 
$\{G^++\Gb^+,\phi_k^{[1]}\}=d\phi_k$. Thus we are reduced to
\begin{equation}
\del_\ib \tilde\Delta_{jk} = \int_0^1 dr \frac{\del}{\del r}\langle\int\phi_\ib^{[1]}
\phi_j(0)\phi_k(r)\rangle_{(0,1)}
\end{equation}
where $\phi_\ib^{[1]}=\frac 12[G^+-\Gb^+,\phi_\ib]$. This is now a sum 
of two boundary terms. When $\phi_k$ hits the boundary at $r=1$, we obtain
a term similar to the Warner term in the supersymmetry variation of
the bulk action \eqref{warner}. By our assumptions, the Warner term
has been canceled as in \eqref{cancelwarner}, in other words 
$\phi_k|_{\del\Sigma}$ is BRST exact on the boundary. So there is no 
contribution from $r=1$. On the other hand, the boundary term at $r=0$, 
when $\phi_j$ and $\phi_k$ collide, can be evaluated by using the bulk 
chiral ring and $tt^*$-fusion. Thus,
\begin{equation}
\del_\ib \tilde\Delta_{jk} = - C_{jkl} g^{\lb l} 
\int_0^1 dr\langle \phi_\lb(0) \phi_\ib^{[1]}(r)\rangle_{(0,1)}
= - C_{jkl} g^{\lb l} \tilde\Delta_{\ib\lb}
\end{equation}
where we have made use of the fact that the angular integration of 
$\phi_\ib^{[1]}$ is trivial once $\phi_j$ and $\phi_k$ have been fused.
This shows that $\tilde\Delta_{ij}$ satisfies exactly the same holomorphic 
anomaly equation as $\Delta_{ij}$. Modulo the holomorphic ambiguity,
this completes our identification of the two-point function on the
disk with the non-holomorphically lifted Griffiths infinitesimal 
invariant of the normal function. We believe that this identification
also holds after the holomorphic ambiguity has been taken into account.
We will be able to verify this in the example from independent information 
in the A-model.

We now have all the machinery in place to extend the holomorphic
anomaly to higher worldsheet topologies.

\subsection{Holomorphic anomaly with D-branes}
\label{openhan}

In analogy to \eqref{Fgclosed}, we want to define the open topological 
amplitude $\F gh$ by an integral over the moduli space $\calm gh$ of 
Riemann surfaces with genus $g$ and $h$ boundary components. In this 
subsection, let us assume $2g+h-2>0$. (We have discussed the disk 
amplitude in the previous subsection, and will return to the annulus 
amplitude in the next.) Given such a Riemann surface $\Sigma_{g,h}$, we 
can close off all the boundaries by gluing in a standard centered disk 
at each boundary component. The data one is forgetting is the length of
the boundary component. This describes $\calm gh$ as a fibration over 
$\calmc g_{h}$, the moduli space of Riemann surfaces of genus $g$ with 
$h$ marked points
\begin{equation}
\eqlabel{fibration}
\bigl(\reals^+\bigr)^h\to\calm gh \to\calmc g_h
\end{equation}
Consequentially, when thinking about the infinitesimal variations of
$\Sigma_{g,h}$, we can isolate those which only change the lengths of
the boundary components, from those which affect also the bulk of the
Riemann surface. We introduce the (real) length moduli by $l^b$, and the
coordinates on $\calmc g_h$ by $m^a$. Let us also denote the Beltrami 
differentials pulled back from $\calmc g_h$ by $\mu_a,\mu_\ab$, $a,\ab=1,
\ldots 3g+h-3$, and the other ones by $\lambda_b$, $b=1,\ldots h$.  

We now define the topological string amplitude $\F gh$ by
\begin{equation}
\eqlabel{Fghopen}
\F gh = \int_{\calm gh}  [dm] [dl] \;\bigl\langle \prod_{a=1}^{3g+h-3} 
\bigl(\int \mu_a G^-\bigr)\bigl(\int \mub_\ab \Gb^-\bigr) 
\prod_{b=1}^{h} \lambda_b(G^-+\Gb^-)\bigr\rangle_{\Sigma_{g,h}}
\end{equation}
It is important here that the $\mu_a$ are complex and can be localized away 
from the boundary $\del \Sigma_{g,h}$. It therefore makes sense to contract 
them with the $G^-,\Gb^-$ individually. On the other hand, the $\lambda_b$ 
are real, and supported near $\del \Sigma_{g,h}$. So we need to contract 
them with the combination that is preserved at the boundary, $G^-+\Gb^-$.

The $\F gh$ are sections of $\call^{2g+h-2}$ over $M$. Recall that
we assume all closed string deformations to be unobstructed by the branes, 
and do not consider any other independent open string moduli, so $M$ is 
the same as before. Also as before, taking a derivative with respect to the 
anti-holomorphic parameter $t^\ib$ brings down the BRST trivial 
operator $\int \phi_\ib^{(2)}=\int d^2z\sqrt{h}\{G^+,[\Gb^+,\phi_\ib]\}$ 
into the correlator. 

When we now pull the action of the BRST operator to the anti-ghosts,
we have to distinguish whether we hit the complex Beltramis from 
$\calmc g_h$ or the real ones corresponding to the variation of the 
lengths of the boundary components. In the latter case, we can only
contract with the BRST charge that is preserved at the boundary, and
remain with an insertion of $\phi_\ib^{[1]}\equiv \frac 12[G^+-\Gb^+,
\phi_\ib]$ in the correlator \cite{bcov2}. Thus, we obtain
\begin{equation}
\eqlabel{rough}
\begin{split}
\del_\ib \F gh &= \int_{\calm gh}[dm][dl] \Biggl[\;
\sum_{a,\ab=1}^{3g+h-3} 4 \frac{\del^2}{\del m^a\del m^\ab} \bigl\langle 
\int \phi_\ib \prod_{\topa{a'\neq a}{\ab'\neq\ab}}
\bigl(\int\mu_{a'} G^-\bigr)
\bigl(\int\mu_{\ab'} \Gb^-\bigr) \\[-.2cm]
&\qquad\qquad\qquad\qquad\qquad\qquad\qquad\qquad\qquad\qquad
\prod_{b=1}^h  \lambda_b(G^-+\Gb^-)
\bigr\rangle_{\Sigma_{g,h}}
+ \\
&\qquad\sum_{b=1}^h 2\frac{d}{dl^b}\bigl\langle \int\phi_\ib^{[1]}
\prod_{a=1}^{3g+h-3} 
\bigl(\int \mu_a G^-\bigr)\bigl(\int \mub_\ab \Gb^-\bigr) 
\prod_{b'\neq b} \lambda_{b'}(G^-+\Gb^-)
\bigr\rangle_{\Sigma_{g,h}} \Biggr]
\end{split}
\end{equation}
(Strictly speaking, there are also terms which mix the base and the fiber
directions of $\calm gh$, but those can be shown to lead to a vanishing
boundary contribution. The arguments are similar to those at the end of 
section 3.1 of BCOV.)

\begin{figure}[ht]
\psfrag{kbar}{$\int \phi_\ib$}
\epsfig{height=5cm,file=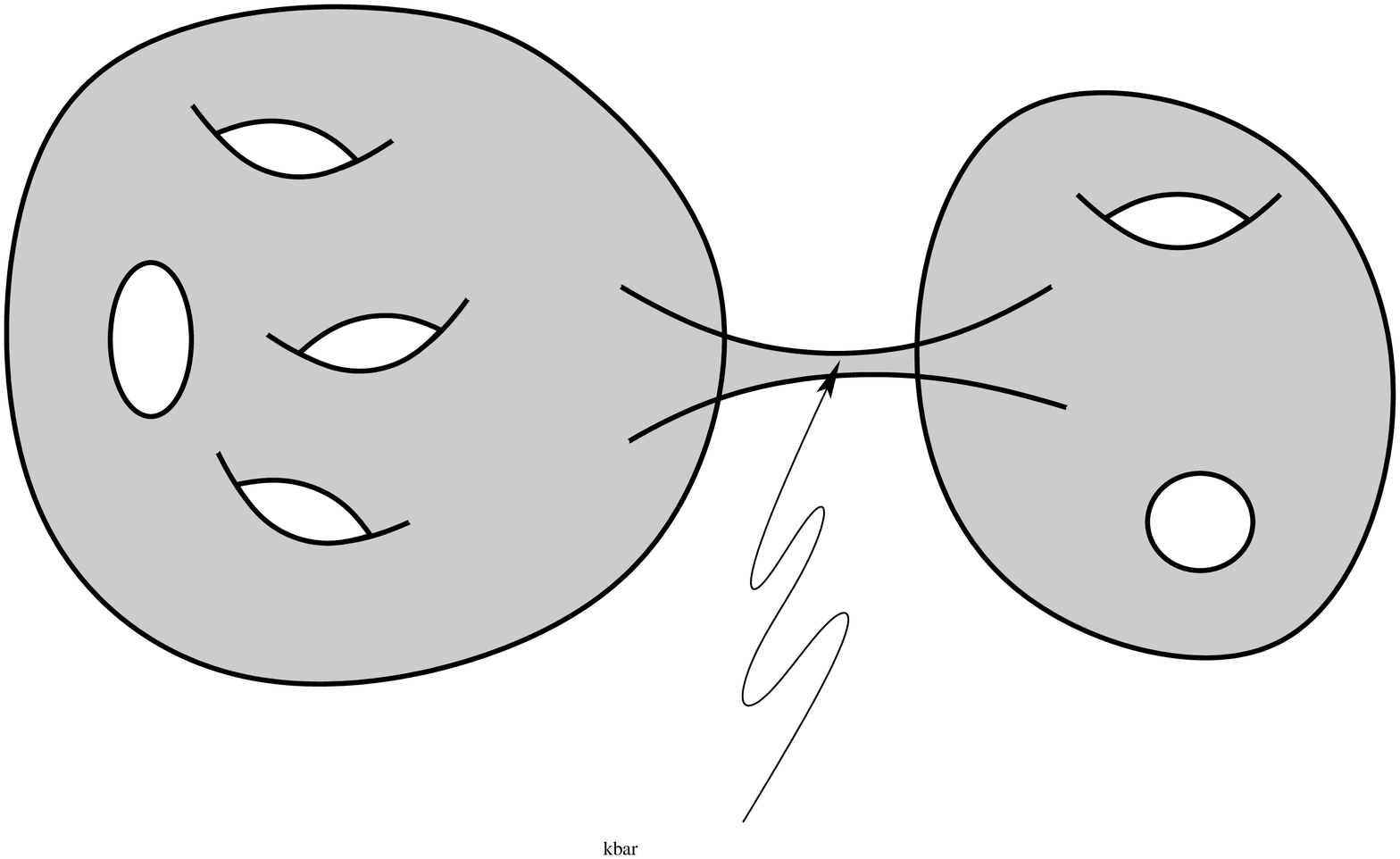}
\qquad
\epsfig{height=5cm,file=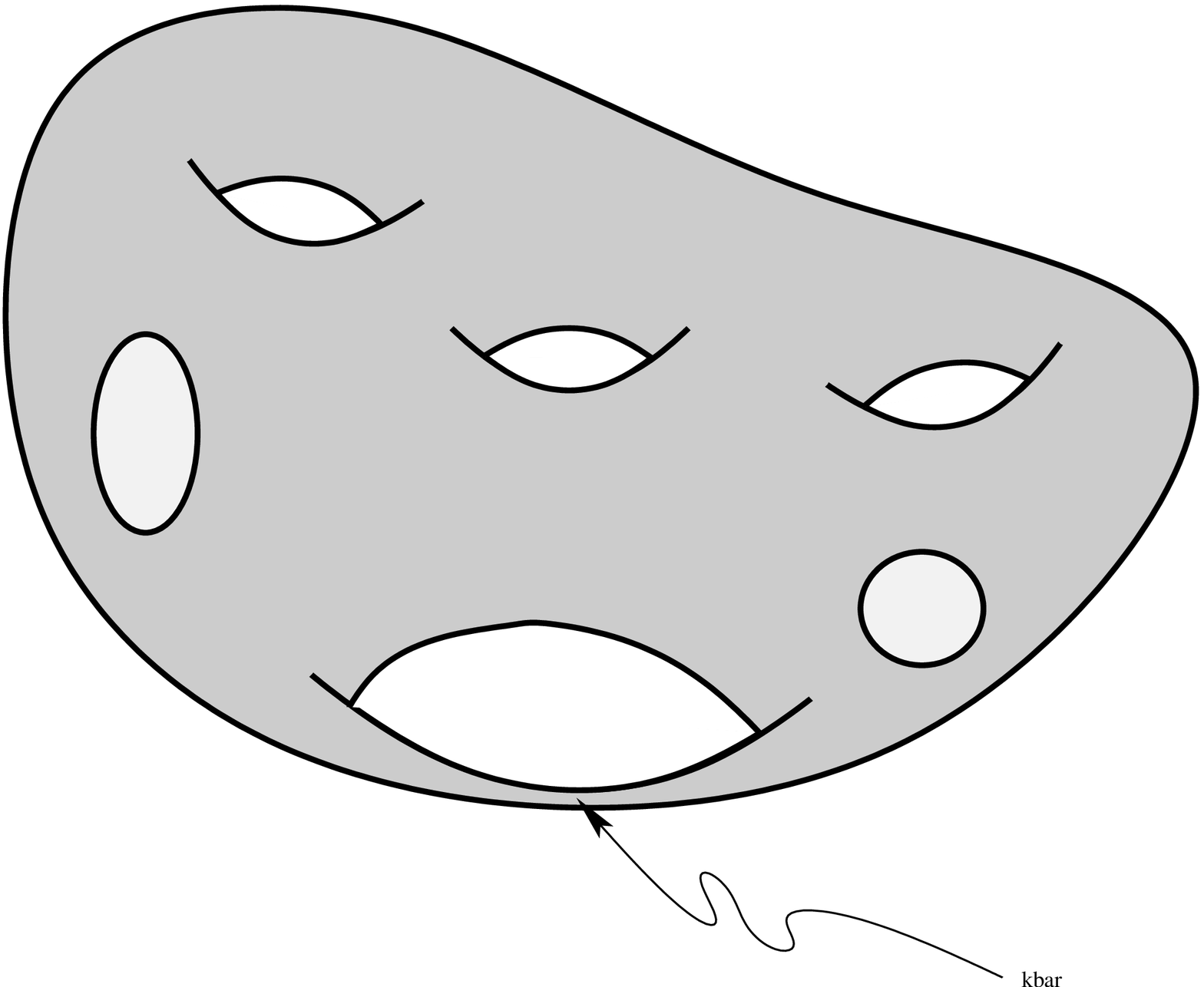}
\caption{When adding boundaries, the two degenerations from Fig.\ 
\ref{degenbcov} remain unaffected.}
\label{degenopen1}
\end{figure}
Now we have to analyze the contribution from the boundary of $\calm gh$. 
As could be expected, the closed string degenerations familiar from 
subsection \ref{cshan} remain essentially unaffected, see Fig.\ 
\ref{degenopen1}. The only difference is that when we split the 
Riemann surface in two pieces, we have to keep track of the distribution 
of the various components of $\del\Sigma_{g,h}$. This leads to a sum 
over $(g_1,h_1)$, $(g_2,h_2)$ with $g_1+g_2=g$, $h_1+h_2=h$. The 
condition to have a stable degeneration imposes the additional
restriction that $3g_i+3h_i/2-2>0$ for $i=1,2$, but in particular 
allows $g_i=0$ as long as $h_i\ge 2$.

There are three types of degenerations which are specific to the
presence of open strings. The first two arise when the Riemann
surface develops a very long strip, which can either split the
Riemann surface in two pieces, or lead to the merging of two
boundary components. It is not hard to see that those degenerations
actually do not contribute. The presence of the very long strip 
projects the intermediate open strings to their ground states, and 
we can therefore replace the strip by the insertion of complete 
sets of chiral boundary fields, $\psi_a$ $\psi_b$, contracted with
the open string topological metric $\eta^{ab}$. But notice that when 
$2g+h-2>0$, at least one of the boundary fields has to be an 
integrated insertion, and as in the closed string case, the only
contribution could have come from marginal ($p=1$) open string
states. Since those are generically absent, we conclude that 
degenerations with long strips do not contribute. 

\begin{figure}[ht]
\psfrag{kbar}{$\int \phi_\ib^{[1]}$}
\epsfig{height=5cm,file=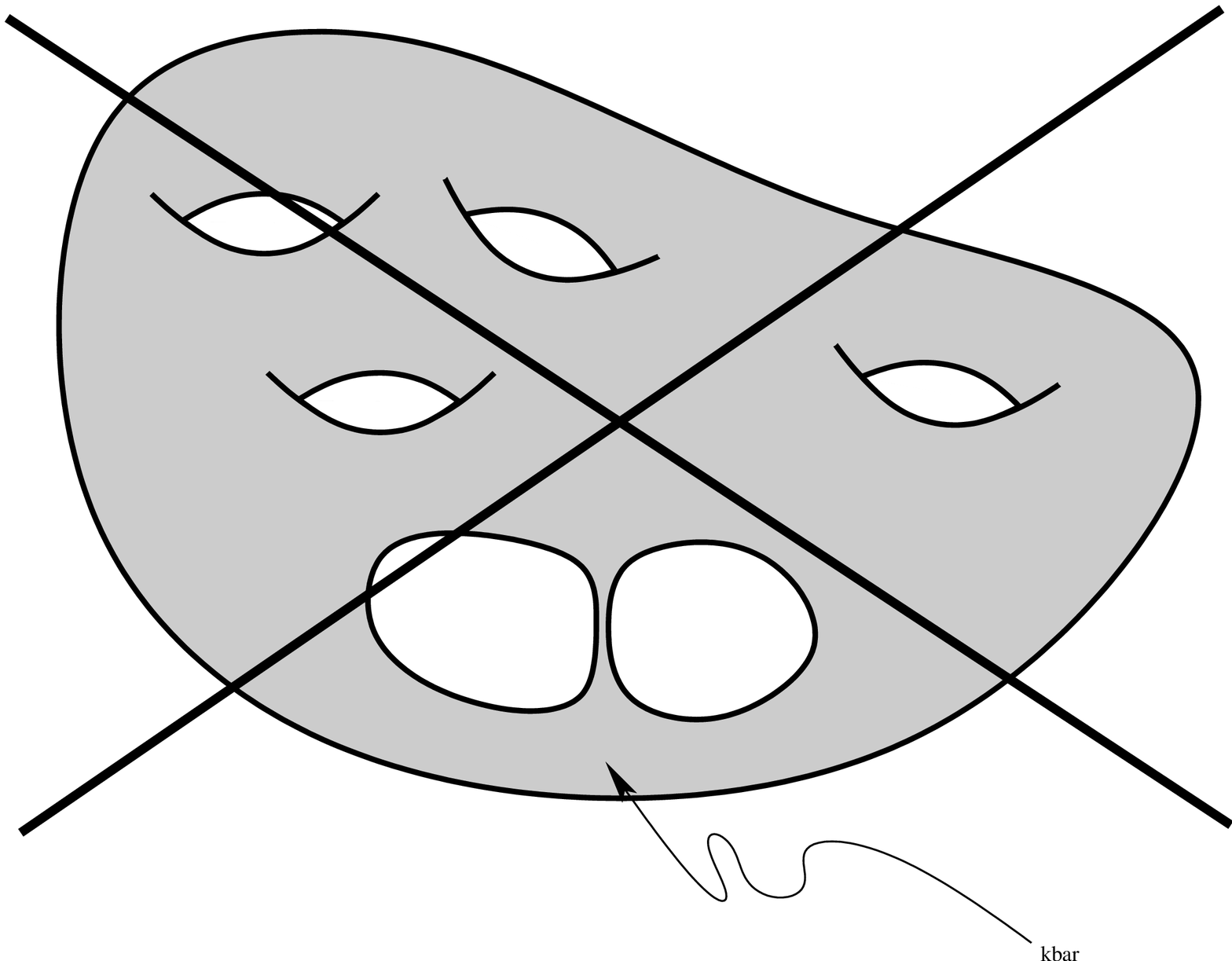}
\qquad
\epsfig{height=5cm,file=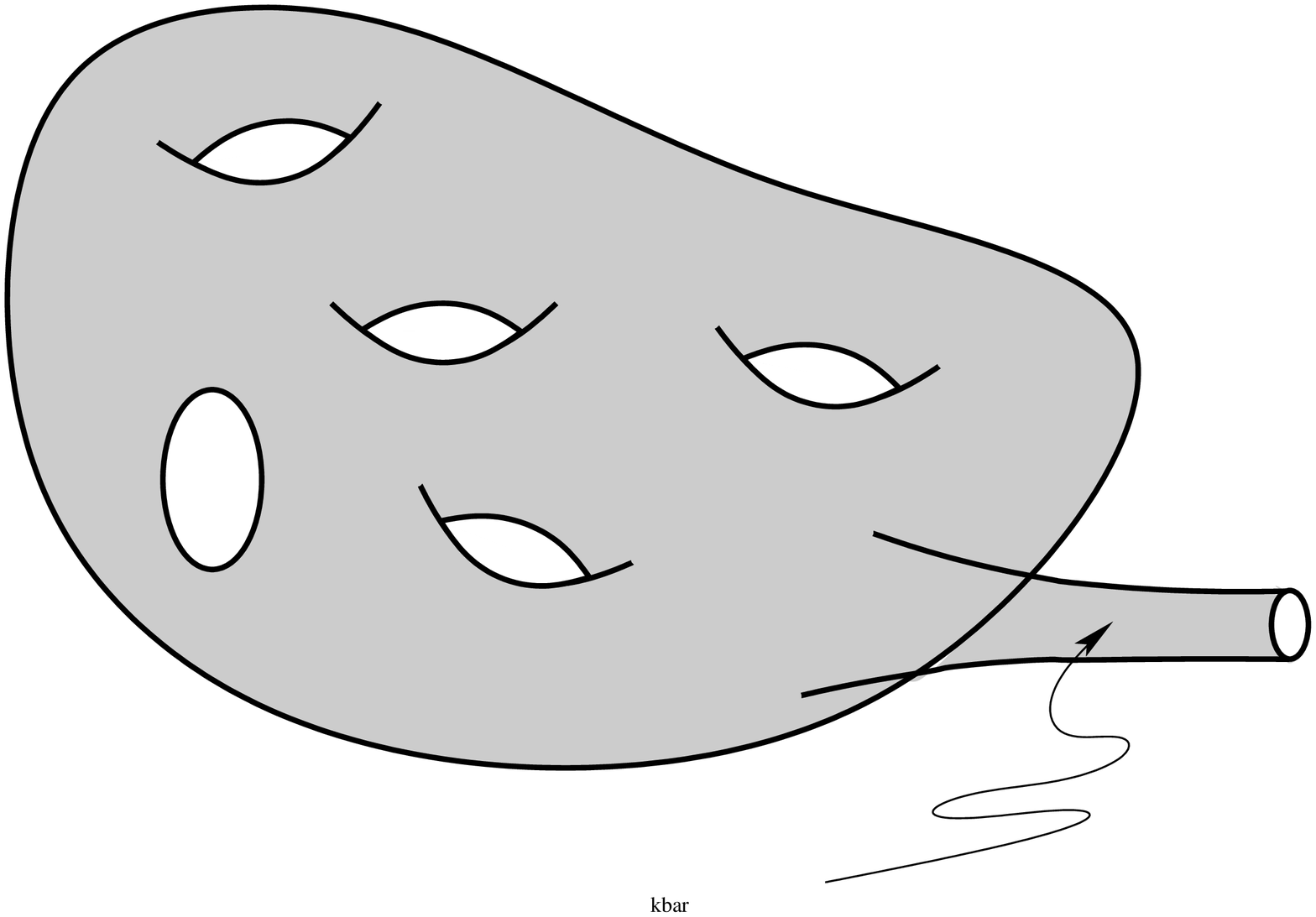}
\caption{The degeneration with an intermediate open string does not 
contribute generically. This leaves only the degeneration in which 
the length of a boundary component shrinks to zero size, or, 
equivalently, the boundary component is separated by a long tube.}
\label{degenopen2}
\end{figure}
The final degeneration we have to take into account arises when
the length of a boundary components shrinks to zero, \ie, $l^b\to 0$
in \eqref{rough}, see Fig.\ \ref{degenopen2}. This is conformally 
equivalent to pulling the boundary very far from the rest of the
Riemann surface via a long tube, at which point it looks more like
a closed string degeneration of the type we have seen before. 
Strictly speaking, however, it would not make sense to pinch off 
the tube because this would have corresponded to a non-stable 
degeneration (in real codimension 2!) involving a disk one-point 
function. Complementarily, we note that {\it as long as the integration 
of the anti-chiral field $\phi_\ib^{[1]}$ is away from the long tube}, 
the intermediate closed string is projected onto the ground states. 
As explained in the previous subsection, our construction is such
that all these one-point functions vanish (by tadpole cancellation 
or Griffiths transversality).

Thus, we only remain with the integration of $\phi_\ib^{[1]}$ over 
the long tube. Note that the angular position of $\phi_\ib^{[1]}$
does not matter after we pull the tube infinitely long. The infinitely
long tube projects the closed string onto their ground states which
can again be represented by inserting a complete set of chiral fields
(only marginal ones contributing). So the rest of the Riemann surface 
has an additional chiral insertion of $\phi_j$, while the long tube 
becomes nothing but the anti-topological disk two-point function,
\begin{equation}
\Delta_{\ib\jb} = \int_{0}^{1} \langle \phi_\ib^{[1]}(r)\phi_\jb(0)
\rangle_{0,1} \,,
\end{equation}
familiar from the previous subsection. The connection to the bulk of
the Riemann surface occurs via the inverse topological metric 
$g^{\jb j}$. This way, we arrive at our final expression for the 
holomorphic anomaly equation in the presence of D-branes,
\begin{equation}
\eqlabel{extended}
\del_\ib \F gh = 
\frac 12 \sum_{\topa{g_1+g_2=g}{h_1+h_2=h}} C_{\ib}^{jk}
\F {g_1}{h_1}_j \F {g_2}{h_2}_k +
\frac 12 C_{\ib}^{jk} \F {g-1}h_{jk} - 
\Delta_{\ib}^j \F g{h-1}_j  \,,
\end{equation}
where we have of course defined
\begin{equation}
\Delta_{\ib}^j \equiv \Delta_{\ib\jb} g^{\jb j} = \Delta_{\ib\jb} 
\ee^K G^{\jb j}
\end{equation}
and the $\F gh$ with subscripts are the amplitudes with closed
string insertions as before. 

We call eq.\ \eqref{extended} the ``extended holomorphic anomaly
equation''. It would be very interesting to clarify in greater 
detail the role played by marginal open string operators in this
equation. As we have mentioned, it seems reasonable to expect 
that actual open string moduli do not enter the $\F gh$ at all.
Massless open string fields with a higher order superpotential 
however will lead to additional singularities at isolated points 
in the moduli space, as we will see in examples in the second half 
of the paper. But before that, let us conclude this first half
by tying up a few loose ends concerning the extended holomorphic
anomaly.

\subsection{Holomorphic anomaly at one loop}
\label{oneloophan}

In our derivation so far, we have factored out the one-loop amplitudes,
on the torus and the annulus for closed and open strings, respectively. 
These amplitudes are somewhat exceptional, because information enters 
which is slightly external to the B-model proper. Nevertheless, they 
fit in the general framework, as we now explain.

The holomorphic anomaly for the closed string one-loop amplitude
was derived in \cite{bcov1}. It takes the form
\begin{equation}
\eqlabel{torushan}
\del_\ib \Fc 1_j = 
\del_\ib\del_j \Fc 1 = \frac 12\tr C_\ib C_j - \frac\chi{24} G_{j\ib}
\end{equation}
Here, negative $\chi$ is the Euler characteristic of the 
Calabi-Yau manifold under consideration. Since 
$\chi=2(h^{21}(Y)-h^{11}(Y))$, the holomorphic anomaly knows 
not only about the vacuum bundle (of rank $2h^{21}+2$), but also 
about the total number of ground states, which is not part of
the special geometry.

In \eqref{torushan}, the second term comes from the collision
of the anti-chiral and the chiral insertion (a special case of
the holomorphic anomaly with insertions, see subsection \ref{insertions}),
while the first comes from the degeneration of the torus to a very 
long tube. It is important to note that the $\tr$ in this first 
term is over the entire vacuum bundle, and not just over the marginal
directions. By using the explicit form of the chiral ring 
multiplication matrices \eqref{struccon}, one finds
\begin{equation}
\del_\ib \Fc 1_j = \frac 12 C_{\ib}^{kl} C_{jkl} - 
\bigl(\frac\chi{24}-1\bigr)G_{j\ib}
\end{equation}
The first term can be viewed as the usual closed string factorization
contribution, while the $-1$ in the second term comes from the propagation 
of the unique ground state of zero charge $(q,\qb)=(0,0)$. The insertion 
of the identity operator leads to a non-trivial contribution in this
case because after factorization, one is dealing with a sphere correlator
with three fixed insertions, and the unintegrated identity operator is 
non-trivial. 

Much the same story holds for the open string as well. The contribution
to the holomorphic anomaly of the annulus diagram from factorization
in the open string channel was in fact already derived by BCOV. It
was found to be
\begin{equation}
\eqlabel{incomplete}
\del_\ib\del_j \F 02 = \del_\ib\del_j \log\det g_{\rm open} + \cdots
\end{equation}
where $g_{\rm open}$ is the $tt^*$-metric on the space of open string
ground states. We have so far been able to neglect the open string
ground states because of the assertion that there are generically
no open string moduli, and non-marginal open string operators do 
not contribute to the geometry of the vacuum bundle or the holomorphic
anomaly for $2g+h-2>0$. For the one-loop amplitudes, however, the open 
string identity operator will also propagate for the same reason as 
in the closed string. 

It is not too hard to determine the $tt^*$-metric on the open string 
ground states of zero charge.\footnote{Discussions with Andrew Neitzke
were essential in clarifying this point, and indeed for this 
entire subsection. I would also like to thank Kentaro Hori for useful 
feedback on the argument.} But before that, let us briefly recall
the description of the open string chiral ring and the topological
metric. Categorically, we can identify the charge $0$ sector of the
open string chiral ring as ${\rm Ext}^0(B,B)$. If $B$ corresponds
to a holomorphic vector bundle $E$, we have the simpler identification
\begin{equation}
\eqlabel{opencc}
{\rm Ext}^0(B,B)\cong H^0({\rm End}E)
\end{equation}
We can think physically of the ${\rm Ext}^0(B,B)$ as the unbroken
generators of the gauge group. Let us also recall that the 
topological metric (Serre pairing) between $H^0({\rm End}E)$
and $H^3({\rm End}E)$ is given (up to a sign) by
\begin{equation}
\langle \psi | \psi' \rangle = \int \tr (\psi\wedge\psi')\wedge\Omega
\,,\qquad \text{for $\psi\in H^0({\rm End E})$, $\psi'\in H^3({\rm End}E)$}
\end{equation}
Let us choose a basis $(f_a)$, $a=1,\ldots N$ of ${\rm Ext}^0(B,B)$,
where we allow for a generic $N=\dim{\rm Ext}^0(B,B)\ge 1$.

To determine the $tt^*$-metric $g_{\rm open}$ in the $p=0$ sector,
it is better to work with the supersymmetric (Ramond) ground states. 
They differ from \eqref{opencc} by spectral flow,
\begin{equation}
\eqlabel{openrgs}
\calh^0 \cong H^0(\sqrt{K_Y}\otimes {\rm End}E)
\end{equation}
where $\sqrt{K_Y}$ is the squareroot of the canonical bundle. So as 
a bundle over $M$, the charge $0$ ground states of the open string 
live in the bundle $\call^{1/2}\otimes {\mathfrak g}$, where 
${\mathfrak g}\cong H^0({\rm End}E)$ is a trivial rank $N$ bundle. 
By the arguments at the beginning of subsection \ref{more}, we do 
not expect any new ambiguities in the open string sector. 
Therefore, the $tt^*$-metric on this space must be
\begin{equation}
(g_{\rm open})_{a\bb} = \langle \Theta f_b|f_a\rangle = \delta_{a\bb} 
\Bigl(\int_Y\Omegab\wedge\Omega\Bigr)^{1/2} = \delta_{a\bb} 
\bigl(g_{0\0b}\bigr)^{1/2}
\end{equation}
where $g_{0\0b}=\ee^{-K}$ is the closed string topological metric
in the zero charge sector \eqref{g00}, and $\delta_{a\bb}$ is a 
constant matrix (independent of the moduli). In particular,
\begin{equation}
\del_\ib\del_j \log\det g_{\rm open} = \frac N2 G_{j\ib}
\end{equation}
where $G_{j\ib}$ is the Weil-Petersson metric on $M$.

Returning to the dots in \eqref{incomplete}, there are two possible
sources for additional contributions. The first comes from the
collision between the chiral and anti-chiral insertion, but this
is easily seen to not contribute. The open counterpart
of the Euler characteristic $\chi(Y)$ is the Witten index 
${\rm tr}(-1)^F$ in the space of $B$--$B$ strings, and this 
vanishes because the intersection pairing is anti-symmetric
for $\hat c=3$.

The final source of contributions to the right hand side of
\eqref{incomplete} comes from factorization in the closed
string channel, in other words from shrinking the inner
boundary of the annulus to zero size. This is the contribution
that is also present in the higher topologies in the previous
subsection. Thus, we arrive at the following holomorphic anomaly 
equation for the annulus amplitude:
\begin{equation}
\eqlabel{annhan}
\del_\ib \F 02_j =
\del_\ib\del_j \F 02 = - \Delta_{jk}\Delta_{\ib}^k + \frac N2 G_{j\ib}
\end{equation}

It was also shown in BCOV that the one-loop topological amplitudes
are given by holomorphic Ray-Singer torsion, and it was argued that 
the holomorphic anomalies at one loop are equivalent to the Quillen 
anomaly. This connection gives a further check on our result 
\eqref{annhan}, although it has to be said that most of the issues
related to Ray-Singer torsion have apparently not been studied for the 
most general objects in the derived category. In the following somewhat
tentative comments, we consider the open string situation, and tacitly 
assume that we are dealing with a holomorphic vector bundle.

In general, the Quillen anomaly gives a formula for the curvature
of the Quillen metric on the (derived) determinant bundle of a
family of hermitian vector bundles $\cale$ over a family of K\"ahler 
manifolds $\caly$. The Quillen metric differs from the Ray-Singer torsion 
by factors of the $L^2$-metric on the cohomology, effectively moving 
the second term in \eqref{annhan} to the LHS of the holomorphic anomaly 
equation. The formula is \cite{bismut}
\begin{equation}
\eqlabel{quillen}
\delbar\del \log \bigl(||\cdot||_{\rm Quillen} \bigr) = 
2\pi\ii \int_Y \td(\caly) \ch(\cale)|_{(1,1)}
\end{equation}
and means that we are to compute the Todd and Chern forms of the
{\it family} with respect to the given metrics on $\cale$ and $\caly$,
integrate over the fiber $Y$ and take the $(1,1)$-piece on the base of
the family. In the case of our interest, the bundles $\cale$ are
typically endomorphism bundles of topologically trivial holomorphic
vector bundles. In this situation, the only contribution to
\eqref{quillen} is expected to come from the algebraic second Chern 
class. This is precisely the quantity that we are computing in terms 
of the normal function and its infinitesimal invariant, as explained 
in subsection \ref{infinvhan}.

This identifies the RHS of \eqref{quillen} with the first
term in \eqref{annhan}. But we can in fact be even more precise about 
this. It appears (see, \eg, \cite{schumacher,fang}) that the curvature 
of the Quillen metric on the determinant bundle can often be interpreted 
as a metric on moduli space of the corresponding geometric objects. The 
holomorphic anomaly of the torus amplitude \eqref{torushan} in this context
is computing simply the Weil-Petersson metric (times $\frac{\chi}{24}$) on 
the complex structure moduli space itself. The moduli space of the 
Calabi-Yau with a B-brane over it can naturally be viewed 
geometrically as the image of the normal function as a section of
the intermediate Jacobian fibration \eqref{intermediate}. It is not 
hard to see that the metric on the normal function that is induced 
from the $tt^*$-metric on the vacuum bundle coincides with the RHS 
of \eqref{annhan}, in precise agreement with the above mentioned
interpretation of the Quillen anomaly.

\subsection{Holomorphic anomaly with insertions}
\label{insertions}

As in the closed string case, we can consider open topological string
amplitudes with insertions of chiral operators in the bulk of the
Riemann surface. These amplitudes are defined by
\begin{equation}
\eqlabel{openins}
\F gh_{i_1,\ldots,i_n} = \int_{\calm gh} \bigl\langle 
\int \phi_{i_1}^{(2)} \cdots \int \phi_{i_n}^{(2)}
\prod_{a=1}^{3g+h-3} 
\bigl(\int \mu_a G^-\bigr)\bigl(\int \mub_\ab \Gb^-\bigr) 
\prod_{b=1}^{h} \lambda_b(G^-+\Gb^-)\bigr\rangle_{\Sigma_{g,h}}
\end{equation}
and can also be obtained from the partition functions by covariant
differentiation
\begin{equation}
\eqlabel{covdiff}
\F gh_{i_1,\ldots,i_{n+1}} = D_{i_{n+1}}\F gh_{i_1,\ldots,i_n}
\end{equation}
The main reason for introducing these amplitudes here is that
they will arise in the next section when we solve the holomorphic
anomaly equation. It is then useful to know the holomorphic
anomaly equations satisfied by these amplitudes with insertions, 
and that the relations in \eqref{covdiff} are consistent with the 
special geometry.

We have, \cf, eq.\ (3.15) in BCOV
\begin{multline}
\eqlabel{inshan}
\del_\ib \F gh_{i_1,\ldots,i_n} = 
\frac 12 \sum_{\topa{g_1+g_2=g}{h_1+h_2=h}} C_{\ib}^{jk}
\sum_{s,\sigma} \frac{1}{s!(n-s)!}
\F {g_1}{h_1}_{ji_{\sigma(1)}\ldots,i_{\sigma(s)}} 
\F {g_2}{h_2}_{ki_{\sigma(s+1)},\ldots,i_{\sigma(n)}} +
\frac 12 C_{\ib}^{jk} \F {g-1}h_{jki_1,\ldots,i_n} \\ -
\Delta_{\ib}^j \F g{h-1}_{ji_1,\ldots,i_n}
-(2g+h-2+n-1)\sum_{s=1}^n G_{i_s\ib}
\F gh_{i_1,\ldots,i_{s-1},i_{s+1},\ldots,i_n}
\end{multline}
The following relations are useful to verify consistency of \eqref{covdiff} 
with the relations of special geometry (namely, the curvature formula 
\eqref{riemann}),
\begin{equation}
\eqlabel{useful}
\begin{split}
D_l C_\ib^{jk} &=0\\
D_l\Delta_\ib^j &= - C_\ib^{jk}\Delta_{kl}
\end{split}
\end{equation}

Finally, we note that the holomorphic anomaly on the disk, \eqref{diskhan} 
can be viewed as a special case of the general holomorphic anomaly 
equation \eqref{inshan}. Thus, just as the holomorphic anomaly for 
higher-point function $n\ge 4$ on the sphere is equivalent to the 
statements of special geometry \cite{bcov2}, we consider \eqref{diskhan} 
as the open string analogue of special geometry.

\subsection{Solution of extended holomorphic anomaly}
\label{hansolutions}

In BCOV, it was shown that the closed string holomorphic
anomaly equation can be solved by a recursive procedure that 
progressively moves the anti-holomorphic derivative to lower and
lower genus amplitudes. The resulting expressions allow a very 
interesting interpretation as Feynman diagrams. However, the 
number of terms quickly grows exponentially with the genus,
and this is not very tractable in practice. More recently,
Yamaguchi and Yau \cite{yayau} have shown that in fact all 
closed topological string amplitudes are polynomial of
a certain degree in a finite number of generators, which is
more pleasant for calculations. It seems almost inevitable 
that a similar statement holds for the extended holomorphic 
anomaly as well. We will however not attempt this here, and 
rather solve the extended holomorphic anomaly in the same way 
as BCOV.

Because of the symmetry of $D_\ib C_{\jb\kb\lb}\in 
{\rm Sym}^4(\bar{T}^*M)\otimes\bar\call^{-2}$, one can locally integrate
\begin{equation}
C_{\ib\jb\kb} = D_\ib D_\jb D_\kb \tilde S
\end{equation}
where $\tilde S\in \bar\call^{-2}$. Namely,
\begin{equation}
C_{\ib}^{jk} = C_{\ib\jb\kb}\ee^{2K} G^{\jb j} G^{\kb k} =
\del_\ib G^{\mb j} \del_\mb G^{\nb k} \del_\nb S
\end{equation}
where $S=\ee^{2K} \tilde S\in \call^{2}$. The relations satisfied
by the quantities
\begin{equation}
\eqlabel{propagators}
S\,,\qquad S^j = G^{\jb j}\del_\jb S\,,\qquad S^{jk} = G^{\jb j}\del_\jb S^k
\end{equation}
play the key role in moving the anti-holomorphic derivative to
the lower genus amplitudes, necessary for solving the holomorphic
anomaly equation. Explicit expressions for the $S$, $S^j$ and $S^{jk}$ 
can be found in BCOV.

We can proceed very similarly in the open string. Since
$D_\ib \Delta_{\jb\kb}\in ({\bar T}^*M)^3\otimes
\bar\call^{-1}$ is symmetric in all three indices, we can write
\begin{equation}
\Delta_{\ib\jb} = D_\ib D_\jb \tilde \Delta
\end{equation}
with $\tilde \Delta\in\bar\call^{-1}$. Then
\begin{equation}
\Delta_\ib^j =  \Delta_{\ib\jb} \ee^K G^{\jb j} = \del_\ib G^{\kb j}
\del_\kb \Delta
\end{equation}
with $\Delta=\ee^{K}\tilde\Delta\in\call$. The key quantities
analogous to \eqref{propagators} are
\begin{equation}
\eqlabel{terminators}
\Delta \, \quad \text{and } \; \Delta^j = G^{\jb j} \del_\jb \Delta
\end{equation}
Explicit expressions for the $\Delta$ and $\Delta^j$ can be obtained
as follows. From the holomorphic anomaly of the disk amplitude,
\begin{equation}
\eqlabel{look}
\del_\ib\Delta_{jk} = - C_{jkl} \Delta_{\ib}^l
\end{equation}
we see that since $C_{jkl}$ is holomorphic,
\begin{equation}
\eqlabel{over}
C_{jkl} \Delta^l =  - \Delta_{jk} + f_{jk}
\end{equation}
where $f_{jk}$ is a holomorphic ambiguity. As in BCOV, we expect
that by a judicious choice of $f_{jk}$, the Yukawa coupling
in \eqref{over} can be inverted, and we can solve for $\Delta^l$,
given $\Delta_{jk}$. If there is only one modulus as in our main
example the quintic, we can set $f_{11}=0$ and obtain
\begin{equation}
\Delta^1 = - \frac{\Delta_{11}}{C_{111}}
\end{equation}
To get an expression for $\Delta$ itself, consider
\begin{align}
\del_\ib D_k \Delta^j &= - C_{\ib}^{jm} C_{kml} \Delta^l
+\delta_k^j  G_{l\ib}\Delta^l - D_k \Delta_\ib^j 
\intertext{
where we have just used the special geometry relation, and
$\del_\ib \Delta^j=\Delta_\ib^j$. Using \eqref{over} 
and the second equation in \eqref{useful}, this is}
&= -C_\ib^{jm} f_{km} + \delta_k^j \del_\ib\Delta
\end{align}
After summing over $j,k$, we get
\begin{equation}
\Delta = \frac 1n\bigl(D_k\Delta^k + S^{mk} f_{mk} +f\bigr)
\end{equation}
where $n={\rm dim}M$, and $f$ is another holomorphic ambiguity.
Let us use these results to solve the extended holomorphic anomaly
equation in some representative examples.

For the annulus, \eqref{annhan} we find
\begin{equation}
\eqlabel{annsolv}
\begin{split}
\del_\ib\del_j\F 02 &= \del_\ib\bigl(-\Delta_{jk}\Delta^k + \frac N2
\del_j K\bigr) - C_{jkl}\Delta_{\ib}^l \Delta^k \\
&=\del_\ib\bigl(-\Delta_{jk}\Delta^k-\frac 12 C_{jkl}\Delta^k\Delta^l
+\frac N2\del_j K\bigr) \\
&= \del_\ib\bigl(-\frac 12(\Delta_{jk}+f_{jk})\Delta^k +\frac N2
\del_j K\bigr)
\end{split}
\end{equation}
where we have used \eqref{over} in the last step. This gives $\F 02$
up to a holomorphic ambiguity.

The next more complicated cases are $(g,h)=(1,1)$ and $(0,3)$.
Let's do $\F 11$.
\begin{equation}
\begin{split}
\del_\ib \F 11 &= \frac 12 C_\ib^{jk} \Delta_{jk} - \F 10_j \Delta_\ib^j \\
&=\del_\ib \bigl(\frac 12 S^{jk}\Delta_{jk} - \F 10_j \Delta^j\bigr)
+\frac 12 S^{jk} C_{jkl} \Delta^l_\ib + \bigl(\frac 12 C_{jkl} C_\ib^{kl} -
\bigl({\frac{\chi}{24}}-1\bigr) G_{j\ib}\bigr)\Delta^j \\
&= \del_\ib\bigl(\frac 12 S^{jk}\Delta_{jk} - \F 10_j\Delta^j
+\frac 12 C_{jkl} S^{kl} \Delta^j -\bigl({\frac{\chi}{24}}-1\bigr)\Delta\bigr)
\end{split}
\end{equation}
For $\F 03$, the result is
\begin{equation}
\F 03 = - \F 02_j\Delta^j + \frac N2\Delta - \frac 12\Delta_{jk}\Delta^j\Delta^k
-\frac 16 C_{jkl}\Delta^j\Delta^k\Delta^l + {\it hol.\ amb.}
\end{equation}
These results admit an interpretation in terms of Feynman graphs
similar to the ones in BCOV.
\vskip -1.2cm
\begin{align}
\epsfig{file=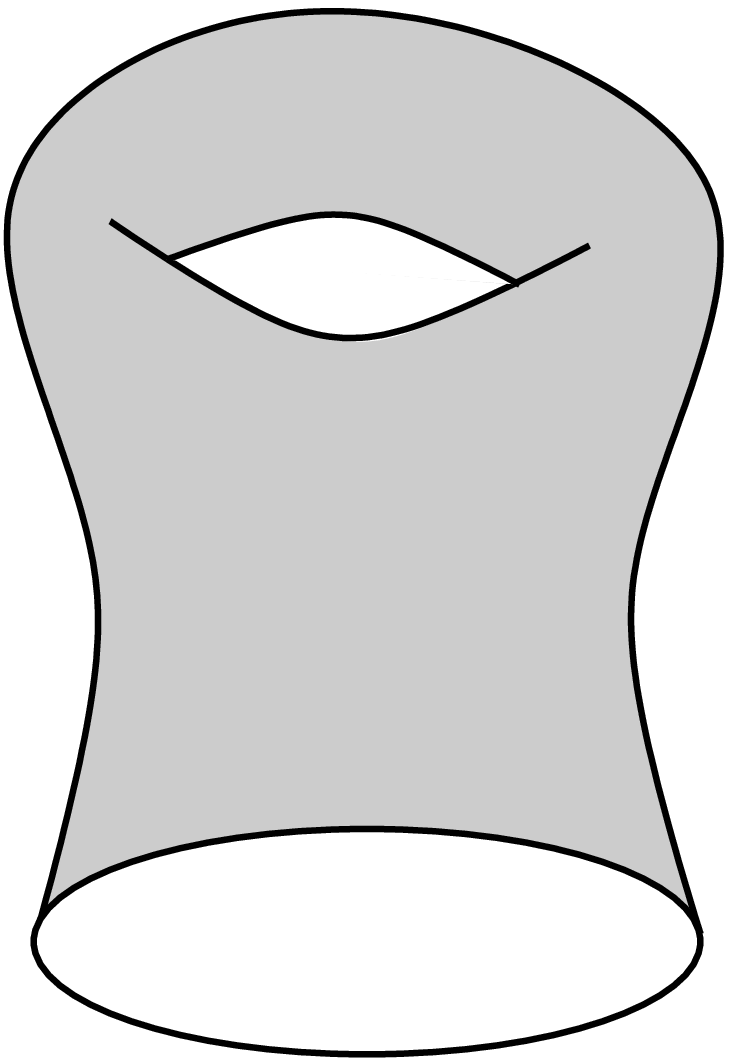,width=2cm} 
&=
\frac 12 \;\;
\epsfig{file=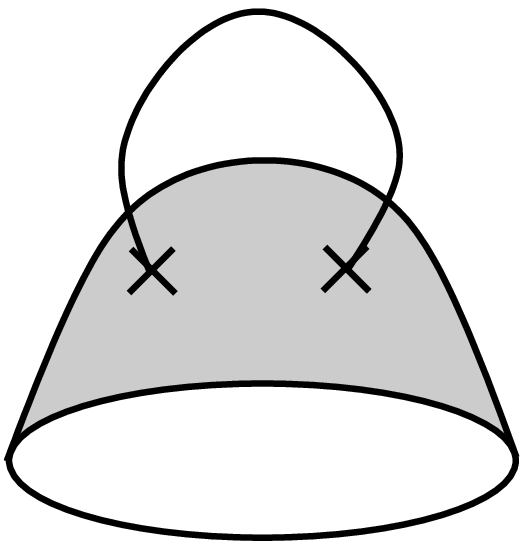,width=1.5cm}
- 
\epsfig{file=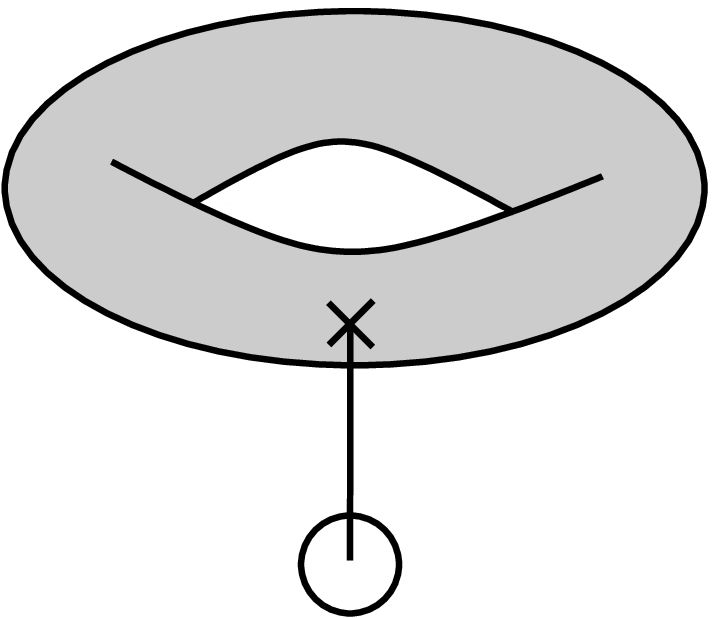,width=1.5cm}
+\frac 12\;\; 
\epsfig{file=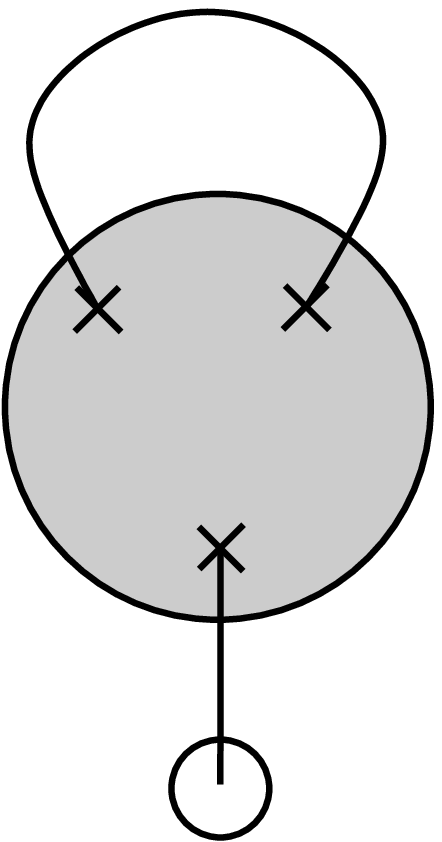,width=1cm}
- 
\epsfig{file=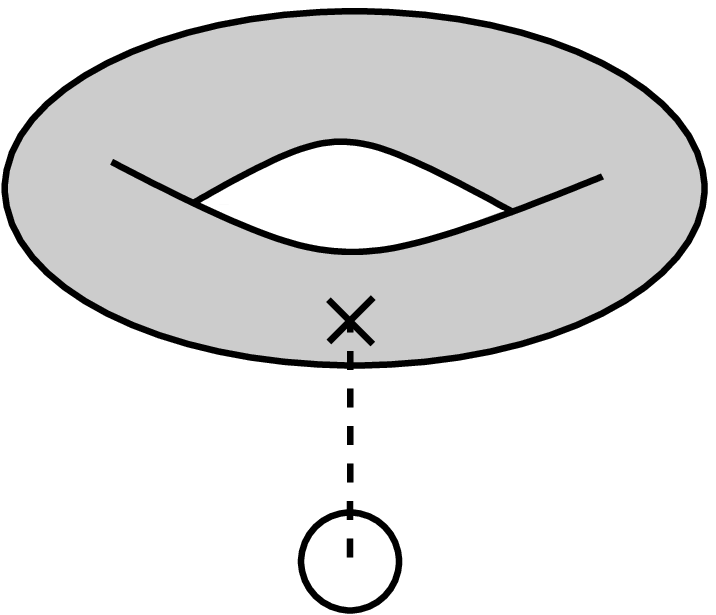,width=1.5cm} 
+{\it hol.\ amb.}\\
\epsfig{file=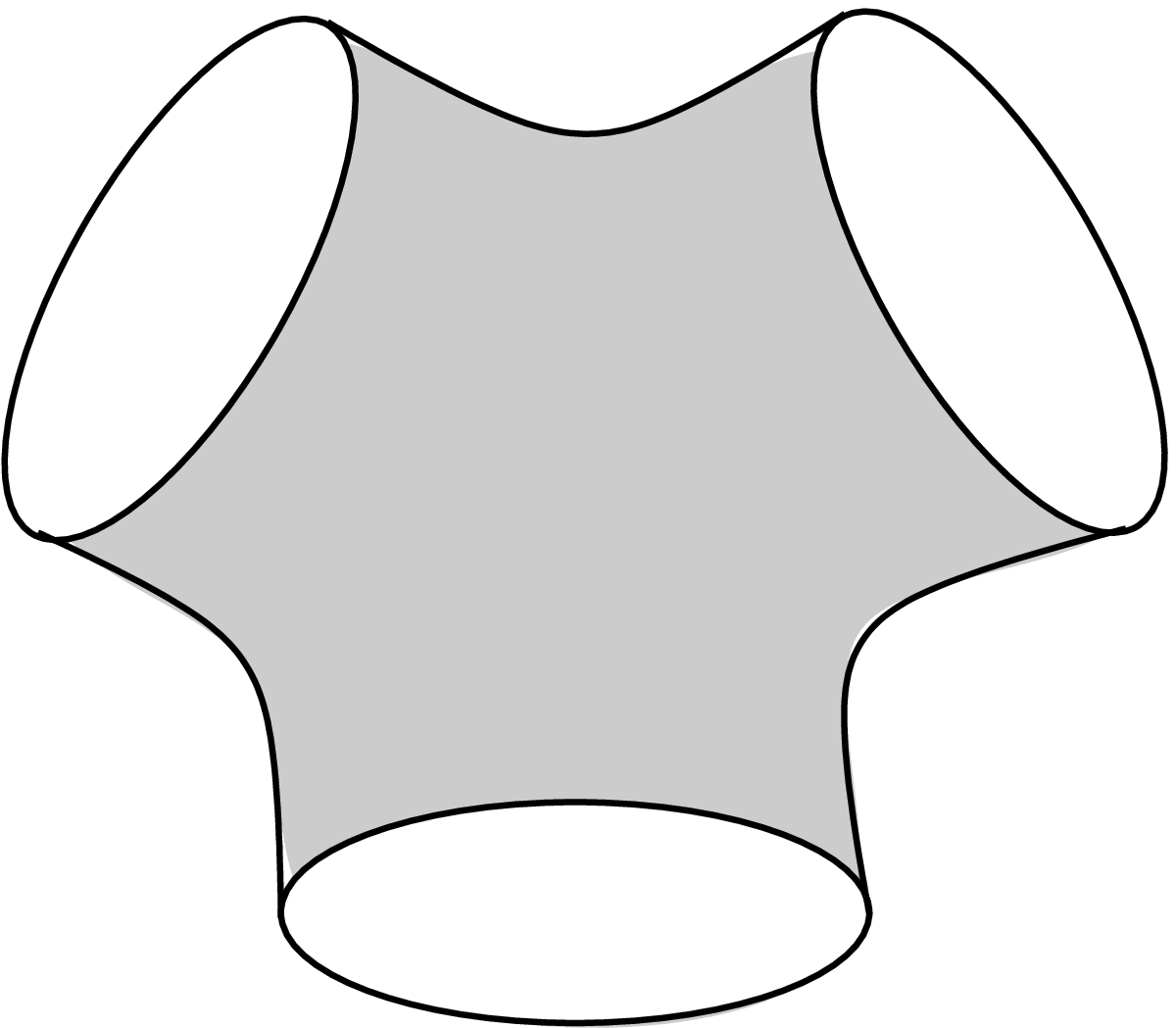,width=2cm}
\;\; &= - 
\epsfig{file=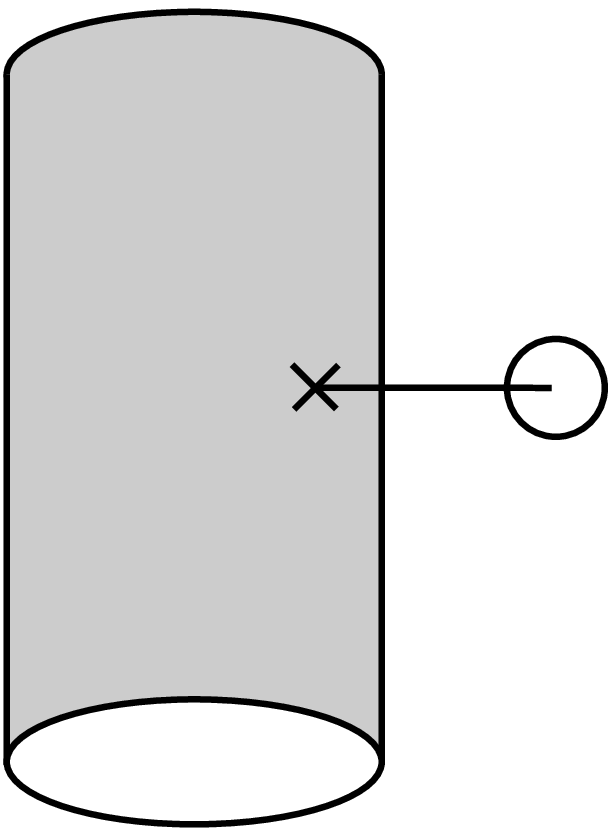,width=1.5cm}
+ 
\epsfig{file=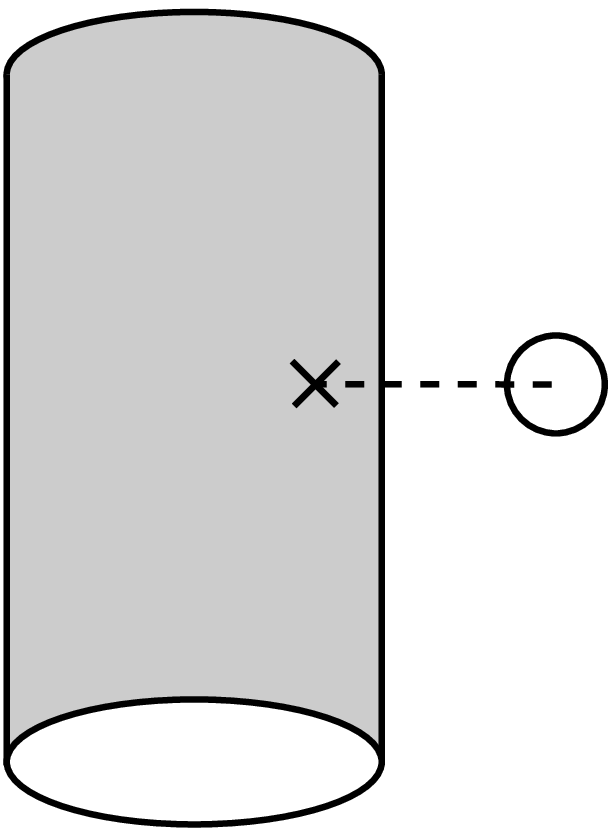,width=1.5cm}
-\frac 12\;\; 
\epsfig{file=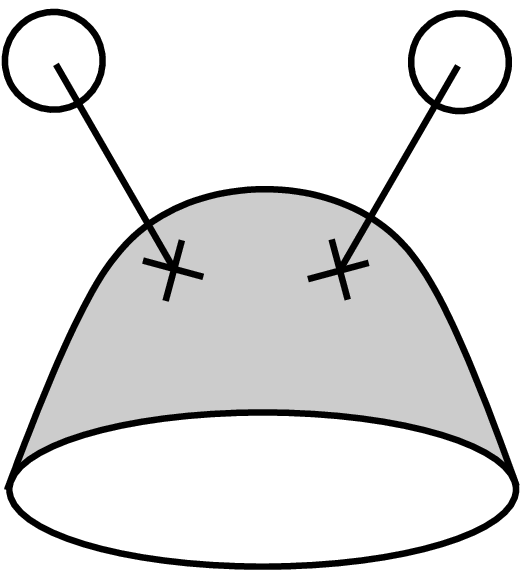,width=1.5cm}
-\frac 16
\epsfig{file=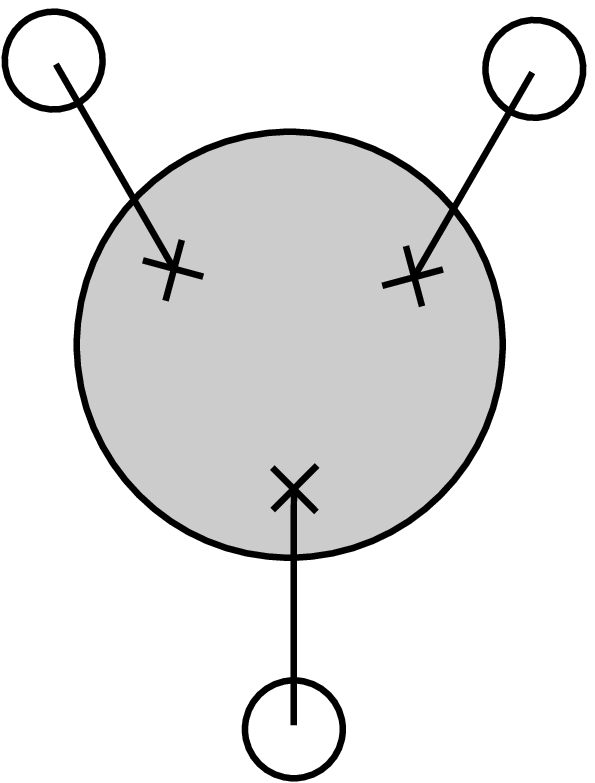,width=1.5cm} 
+{\it hol.\ amb.}
\end{align}
\vskip .5cm
\noindent where
\begin{equation}
\epsfig{file=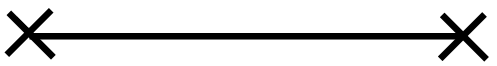,width=1cm} 
= S^{ij} \,,
\qquad
\epsfig{file=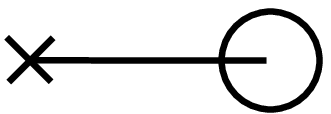,width=1cm} 
= \Delta^i \,,
\qquad
\epsfig{file=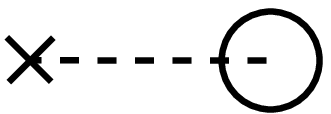,width=1cm} 
= \Delta
\end{equation}
and all other conventions are as in BCOV. It should not be hard to show that 
such a graphical expansion is valid for all, $(g,h)$.

\section{Open Topological String Amplitudes on the Real Quintic}

\begin{equation}
X := \{ P(z) = 0 \} \subset \projective^4
\end{equation}
where $P$ is a homogeneous polynomial of degree $5$ in $5$ variables
$z_1,\ldots, z_5$. Our interest is in the A-model on $X$, which depends 
on the complexified K\"ahler parameter $t$ of $X$. Assume that $X$ is 
{\it real} in the sense that all coefficients of $P$ are real (or have 
the same phase). Then the real locus 
\begin{equation}
L = \{ z_i = \bar z_i \} \subset X
\end{equation}
is a Lagrangian submanifold. If $X$ is Fermat, $P(z) = \sum z_i^5$, 
we find $L\cong \reals\projective^3$. To wrap an A-brane on $L$, we
need to specify a flat gauge field, for which there are two choices
because $H_1(L;\zet)=\zet_2$. The BPS domainwalls between the
corresponding worldvolume vacua are classified by those classes in 
$H_2(X;L)\cong\zet$ with non-trivial image in $H_1(L;\zet)$.
Modulo $H_2(X;\zet)$, the tension of those domainwalls is for large 
$t$ given by \cite{open}
\begin{equation}
\eqlabel{funda}
\calt(t) = \frac t2 \pm \bigl(\frac 14 + \frac{15}{\pi^2} q^{1/2} + 
\cdots\bigr)
\end{equation}
where $q\equiv \ee^{2\pi\ii t}$ and the dots, which are of higher order 
in $q$, correspond to corrections from worldsheet (disk) instantons.

The domainwall tension $\calt(t)$ effectively computes the topological 
string amplitudes on the disk, as we have explained in the previous
section. More precisely, the relation of $\calt(t)$ to topological
amplitudes is similar to the relation between the genus $0$ prepotential
$\Fc 0$ and topological amplitudes on the sphere, which can be obtained 
by differentiation. We will here compute topological string amplitudes 
for higher worldsheet topologies, with boundary on the real quintic,
by using the extended holomorphic anomaly equation and the tree-level 
data \eqref{funda}.

\subsection{Conventions}

There is also a B-model description of the above situation.\footnote{Our
conventions follow, with minor changes, those of \cite{cdgp,bcov2}, and 
later \cite{hkq}.} The mirror quintic $Y$ is obtained from the one-parameter 
family of quintics
\begin{equation}
\{W=x_1^5+x_2^5+x_3^5+x_4^5+x_5^5 - 5\psi x_1x_2x_3x_4x_5 = 0\}\subset
\projective^4
\end{equation}
after a $(\zet_5)^3$ quotient. The B-model can be equivalently described 
as a Landau-Ginzburg model with worldsheet superpotential $W$. The
mirror of $L$ with its two vacua is a twin family of matrix 
factorizations of $W$, described in \cite{howa,mowa}.

For the appropriate choice of holomorphic three-form, the Picard-Fuchs 
operator for the family $Y$ is
\begin{equation}
\call = \theta^4 - 5 z (5\theta+1)(5\theta+2)(5\theta+3)(5\theta+4)
\end{equation}
where $z=(5\psi)^{-5}$, and $\theta=zd/dz$. As shown in \cite{open,psw,mowa}, 
the domainwall tension satisfies the inhomogeneous Picard-Fuchs equation 
\begin{equation}
\eqlabel{inhomo}
\call \calt(z) = \frac{15}{16\pi^2} \sqrt{z}
\end{equation}
This equation of course determines $\calt$ only up to a solution of
the homogeneous Picard-Fuchs equation. This is more ambiguity than
we can tolerate when we lift the normal function from the intermediate
Jacobian $J^3(Y)=H^3(Y;\complex)/ (F^2H^3(Y)+H^3(Y;\zet))$ to $H^3(Y;\reals)
\subset H^3(Y;\complex)$, see subsection \ref{infinvhan}. Indeed, not 
any arbitrary complex linear combination of solutions of the Picard-Fuchs 
equation can be obtained by integrating $\Omega$ over an actual real 
three-cycle of $Y$. However note that the disk information enters the 
extended holomorphic anomaly via the infinitesimal invariant 
\eqref{infinv}, which is invariant under adding to $\calt$ a solution
of the homogeneous Picard-Fuchs equation corresponding to an arbitrary
{\it real} (but not necessarily integral) combination of periods. 
In any case, we know from \eqref{funda} exactly which solution of
\eqref{inhomo} we want around large volume, and by analytical continuation
we will know the reality properties of $\calt(z)$ also around other
points in moduli space.

When analyzing the physical interpretation of topological string 
amplitudes in various regions of moduli space, one should do so by 
expanding in the so-called canonical coordinates appropriate for the 
region of interest. In this ``holomorphic limit'', the $\F gh$ 
become holomorphic function of the canonical coordinates, which is
indeed required for interpreting the amplitudes in terms of a
4-dimensional effective action. 

Consider a point $m\in M$ defined by some local coordinates $z=\zb=0$.
Canonical coordinates and an accompanying canonical gauge for the 
holomorphic three-form \cite{bcov2} are defined by the property that
the connection coefficients and all their holomorphic derivatives
vanish to order $O(\zb)$ around $m$. Canonical coordinates always
exist and are the K\"ahler analogue of geodesic normal coordinates
in Riemannian geometry. (Around singular points in $M$, some care is
required to remove the most divergent terms, but this can always be 
done.)

To take the holomorphic limit of open topological string amplitudes,
we proceed as follows. We note again that the infinitesimal invariant
\eqref{infinv} is invariant under modifying $\calt$ by a {\it real}
linear combination of periods. If $n={\rm dim}M$, there are $(2n+2)$
real periods, $\varpi_i$, $i=1,\ldots ,2n+2$. Around a generic point 
in $M$, this freedom is enough to find a real linear combination
\begin{equation}
\tilde\calt = \calt + \alpha^i\varpi_i \qquad \text{$\alpha^i$ real}
\end{equation}
such that $\tilde\calt$ together with all its first holomorphic 
derivatives vanish to first order at $m$. Therefore, 
$D_\kb\overline{\tilde\calt}=O(\zb)$ and the holomorphic limit
of \eqref{infinv} is simply
\begin{equation}
\eqlabel{simply}
\lim_{\zb\to 0} \Delta_{ij} =\lim_{\zb\to 0} D_i D_j \tilde\calt 
= \del_i\del_j \tilde\calt
\end{equation}
where the latter equality holds with respect to canonical coordinates
and three-form gauge. This is how we will do our calculations below.

Let's summarize what we know already. The Yukawa coupling on the quintic 
is in the Candelas gauge given by
\begin{equation}
C_{\psi\psi\psi} = \frac{5^4\psi^2}{1-\psi^5}
\end{equation}
and the Euler character is
\begin{equation}
\chi\equiv \chi(X)=-200
\end{equation}
We can use the same propagators as in BCOV, 
\begin{align}
\notag
S^{\psi\psi} &= \frac{1}{C_{\psi\psi\psi}} 
\del_\psi \log \bigl(G^{\bar\psi\psi} (\psi\ee^K)^2\bigr) \\
\eqlabel{exprop} 
S^{\psi} &= \frac{1}{C_{\psi\psi\psi}} \bigl[
\bigl(\del_\psi\log ( \psi \ee^K) \bigr)^2 - D_\psi\del_\psi\log (\psi \ee^K)\bigr] \\
\notag
S &=\bigl[S^\psi-\frac 12 D_\psi S^{\psi\psi} - 
\frac 12 (S^{\psi\psi})^2C_{\psi\psi\psi}\bigr]\del_\psi\log ( \psi \ee^K)
+ \frac 12 D_\psi S^\psi +\frac 12 S^{\psi\psi} S^{\psi} C_{\psi\psi\psi}
\end{align}
while our terminators are given by the expressions in subsection 
\ref{hansolutions}
\begin{equation}
\eqlabel{exterm}
\begin{split}
\Delta^\psi &= -\frac{\Delta_{\psi\psi}}{C_{\psi\psi\psi}} \\
\Delta & = D_\psi\Delta^\psi
\end{split}
\end{equation}
The solutions of the extended holomorphic anomaly for low $(g,h)$ are,
\begin{equation}
\eqlabel{expsols}
\begin{split}
\F 02_\psi & = -\Delta_{\psi\psi} \Delta^\psi + \frac 12 \del_\psi K + \f 02_\psi \\
\F 11 & = -\F 10_\psi\Delta^\psi -\bigl(\frac{\chi}{24}-1\bigr) \Delta + \f 11 \\
\F 03 & = -\F 02_\psi\Delta^\psi +\frac 12 \Delta - 
\frac 13\Delta_{\psi\psi}\Delta^\psi\Delta^\psi + \f 03\\
\end{split}
\end{equation}
Here $\f 02_\psi$, $\f 11$, and $\f 03$ are holomorphic ambiguities which for 
reasons explained in more detail below, we parametrize as follows
\begin{equation}
\eqlabel{expambi}
\begin{split}
\f 02_\psi &= \A 020 \del_\psi \log(\psi^{-5}-1) \\
\f 11 &=  \sqrt 5\A 110 \psi^{-5/2} + \sqrt 5\A 111 \frac{\psi^{5/2}}{1-\psi^5}\\
\f 03 &= \sqrt 5\A 030 \psi^{-5/2} + \sqrt 5\A 031 \frac{\psi^{5/2}}{1-\psi^5}
\end{split}
\end{equation}

\subsection{Large volume expansion}

The large complex structure point $\psi\to\infty$, $z\to 0$ is a point
of maximal unipotent monodromy and the most convenient for finding an 
integral basis of periods. Such a basis is determined by \cite{cdgp},
\begin{equation}
\eqlabel{gauge}
\begin{split}
X^0 = \varpi_0 (z) = \sum_{m=0}^\infty &\frac{(5m)!}{(m!)^5} z^m \\
X^1 = \varpi_1 (z) = \frac 1{2\pi\ii  }&\Bigl[\varpi_0(z) \log z + 5 \sum_{m=1}^\infty 
\frac{(5m)!}{(m!)^5} z^m \bigl[\Psi(1+5m)-\Psi(1+m)\bigr]\Bigl] \\
F_1 = -\frac{5}{2(2\pi\ii)^2} \Bigl[\varpi_0& \log^2 z+(1540z+1620450z^2+\cdots)\log z\\
&\qquad\qquad
+1150 z + \frac{4208174}2 z^2+\cdots \Bigr] + \frac{25}{12}\varpi_0 -
\frac{11}{2} \varpi_1  \\
F_0 = \frac{5}{6(2\pi\ii)^3} \Bigl[\varpi_0& \log^3 z+ (2310z+2430675 z^2+\cdots)
\log^2 z
+ (3450z+\\ \frac{12624525}2 z^2+\cdots)\log& z -  6900z-\frac{9895125}2z^2+\cdots
\Bigr] 
+ \frac{25}{12} \varpi_1 - \frac{25\ii\zeta(3)}{\pi^3}\varpi_0  \\
\end{split}
\end{equation}
The canonical coordinate at large complex structure is the special
coordinate 
\begin{equation}
t = \frac{\varpi_1}{\varpi_0}
\end{equation}
and taking the holomorphic limit amounts to putting 
\begin{equation}
\ee^{-K} = \varpi_0\,,  \qquad\qquad
G_{\psi\bar\psi} = 2\pi\ii \frac{dt}{d\psi}
\end{equation}
in the formulas like \eqref{exprop} and \eqref{expsols}. The prepotential of the 
quintic is in this gauge and coordinate
\begin{equation}
\Fc 0 = (2\pi\ii)^3 \Bigl[ -\frac 56t^3 -\frac{11}4t^2 + \frac{25}{12}t
-\frac{25\ii\zeta(3)}{2\pi^3} + \frac{1}{(2\pi\ii)^3} \bigl(2875 q + 
\frac{4876875}8 q^2 +\cdots\bigr) \Bigr]
\end{equation}
where $q\equiv \ee^{2\pi\ii t}$. Turning to the real quintic, we have 
found the domainwall tension to be the solution of \eqref{inhomo} with
asymptotics
\begin{equation}
\eqlabel{raw}
\frac{\varpi_1}2 - \frac{\varpi_0}4 - \frac{15}{\pi^2}
\sum_{m=0}^\infty \frac{\Gamma(7/2+5m)}{\Gamma(7/2)}
\frac{\Gamma(3/2)^5}{\Gamma(3/2+m)^5}\, z^{m+1/2} 
\end{equation}
The corresponding A-model expansion is
\begin{equation}
\calt = (2\pi\ii)^2 \Bigl[ \frac{t}2 - \frac 14 + \frac{2}{(2\pi\ii)^2}
\bigl(30 q^{1/2} + \frac{4600}3 q^{3/2}+ \cdots \bigr)\Bigr]
\end{equation}
Note that taking the holomorphic limit of the infinitesimal invariant is
particularly simple at large volume. Since \eqref{raw} differs from a 
real (although not integral) period only in exponentially small instanton
corrections, we can compute $\Delta_{\psi\psi}$ in the holomorphic limit 
(\cf, \eqref{simply}) simply by forgetting the $\varpi_0$ and $\varpi_1$ 
contribution in \eqref{raw}. We find
\begin{equation}
\Delta_{tt} = \lim_{\bar\psi\to\infty} 
\ee^K (G_{\psi\bar\psi})^{-2}\Delta_{\psi\psi} = 
15 q^{1/2} + 6900 q^{3/2} + 13603140 q^{5/2} + \cdots 
\end{equation}
and can now plug all this data into \eqref{expsols}.

To fix the holomorphic ambiguity, we can make use of the enumerative 
interpretation of the topological string amplitudes in terms of
BPS invariants \cite{gova,oova,lmv}. To adapt the general multicover
formula from \cite{lmv} to our situation, we have to take into account 
that the only one-cycle on $L$ by which we can classify the boundary
data of BPS invariants is a torsion cycle. A most natural conjecture is
that when we expand the $\F gh$ as
\begin{equation}
\eqlabel{lmv}
\sum_{g=0}^\infty g_s^{2g+h-2} \F gh = (-1)^{h-1}
\sum_{g=0}^\infty \sum_{\topa{d\equiv h\bmod 2}{k\;{\rm odd}}} 
\n gh{{d}}
\frac 1k\Bigl(2\sin \frac{kg_s}2\Bigl)^{2g+h-2} q^{kd/2} 
\end{equation}
all $\n ghd$ should be integer. Notice that this multicover formula
(as those in \cite{lmv}) does not mix worldsheets with different numbers 
of boundaries. Also note that we have neglected any constant map 
contributions which would show up at $d=0$. We expect that there are 
such contributions only at $(g,h)=(0,1)$, where $\n 010=\frac 12$ (see 
\cite{open}), and $(g,h)=(0,2)$, where $\n 020$ is given by ordinary 
(Reidemeister, or analytic Ray-Singer) torsion of $L$. Both contributions 
drop out of the higher-loop computations. The statement that the expansion 
\eqref{lmv} is regular means that the holomorphic ambiguities 
should all vanish as $\psi\to\infty$. This is satisfied by our 
ansatz \eqref{expambi}.

\medskip

\noindent {\bf A correction}

\smallskip

\noindent It has recently become clear \cite{topor} that an ansatz of the form \eqref{lmv} is 
probably too optimistic. Let us briefly summarize the main objection in the
current context. Naively, one would attempt to identify the
$\n ghd$ as BPS invariants enumerating oriented Riemann surfaces of 
genus $g$ with $h$ boundary components in the class $d\in H_2(X,L;\zet)
\cong\zet$, such that each boundary component is mapped to the non-trivial 
class of $H_1(L;\zet)\cong\zet_2$. At the level of Gromov-Witten invariants,
it should indeed be possible to distinguish different topologically non-trivial
boundary components by coupling to Wilson line observables on an appropriate
stack of D-branes. When some boundary components are trivial in $H_1(L;\zet)$, 
one expects a mixing with unoriented Riemann surfaces. The physics definition of
the integral invariants, however, makes explicit reference to a certain 
supersymmetric string background. As a consequence, it appears that {\it when 
$H_1(L;\zet)$ is torsion} (or, more generally, when open string vacua are discrete)
one can extract integral invariants from the topological string amplitudes
only when unoriented worldsheets are included from the beginning and the number
of D-branes is fixed, thus precluding the measurement of indivual $\n ghd$. See 
\cite{topor} for a complete discussion of these issues, as well as the
relationship to real enumerative invariants.

In retrospect, it is somewhat surprising that with the naive ansatz \eqref{lmv},
the ambiguities parameterized in \eqref{expambi} can in fact be fixed in such a 
way that the expansion coefficients are nevertheless integer, and moreover
satisfy the low-degree vanishing relation that $\n ghd$ must vanish whenever 
$\n {2g+h-1}{0}{{d}}$ does. This constraint arises from the observation that
since our Lagrangian $L$ is defined as the fixed point set of an anti-holomorphic 
involution, we can by complex conjugation complement any curve of genus $g$ 
with $h$ boundaries on $L$ to a holomorphic curve of genus $2g+h-1$ with no 
boundaries. If some given curve contributes to $\n ghd$, the doubled 
curve would have to contribute to $\n{2g+h-1}{0}{{d}}$. We have found that for 
$(g,h)=(0,2)$, $(1,1)$ and $(0,3)$, these conditions are sufficient to 
completely determine the holomorphic ambiguity, and the rest of the expansion 
\eqref{lmv} is then integral. The holomorphic ambiguities in \eqref{expambi} 
take the values
\begin{equation}
\eqlabel{values}
\begin{split}
\A 020 =  - &\frac 3{250}\,,\qquad
\A 110 = \frac{211}{1250} \,,\qquad
\A 111 = \frac{9}{5000}
\\
&\A 030 = \frac{1887}{312500}\,,\qquad
\A 031 = \frac{3}{78125}
\end{split}
\end{equation}
but we refrain from presenting the explicit results for the $\n ghd$.

We now proceed to the expansion of the topological amplitudes
around the other special points in the moduli space, first the 
Gepner point, and then the conifold.

\subsection{Orbifold expansion}

A basis of solutions of the homogeneous Picard-Fuchs equation
$\call\varpi=0$ around the Gepner point $\psi=0$ is given by 
($k=1,2,3,4$)
\begin{equation}
\pio_k = \psi^k \sum_{m=0}^\infty \frac{\Gamma(k/5+m)^5}{\Gamma(k/5)^5}
\frac{\Gamma(k)}{\Gamma(k+5m)} (5\psi)^{5m}
\end{equation}
This basis is neither integral nor real, but the relation to the integral 
basis at large volume \eqref{gauge} is well-understood. We will not need
the details here. A particular solution of the inhomogeneous Picard-Fuchs 
is
\begin{equation}
\tauo = -\frac{4}{3}\sum_{m=0}^\infty \frac{\Gamma(-3/2-5m)}{\Gamma(-3/2)}
\frac{\Gamma(1/2)^5}{\Gamma(1/2-m)^5} (5\psi)^{5(m+1/2)}
\end{equation}

We now come to an important point. In parameterizing the holomorphic ambiguity
\eqref{expambi}, we have allowed for singularities around the Gepner point,
whereas all closed string amplitudes are regular there. This is due to an
important property of open strings ending on the real quintic that we have
already mentioned in the introduction. Recall that when we start out at
large volume, the brane wrapped on the real quintic has two vacua,
corresponding to the choice of a discrete Wilson line. Based on B-model
considerations, it was shown in \cite{howa} that those two vacua coalesce 
as we approach the point $\psi=0$ in K\"ahler moduli space. The vacuum 
structure can be described locally around $\psi=0$ by a superpotential 
\cite{bdlr}
\begin{equation}
\calw = \psi\varphi - \frac 13 \varphi^3
\end{equation}
where $\varphi$ is the open string field that becomes massless at $\psi=0$.
This is just as in \eqref{generically} with $n=2$. This behavior should
be accompanied by the appearance of a tensionless domainwall in the
BPS spectrum of the 4d theory. Indeed, it was shown in \cite{open} that
after analytic continuation of \eqref{raw}, $\calt-\tauo$ is an {\it 
integral} period. Since $\tauo\sim \psi^{5/2}$ vanishes faster 
than the two lightest periods $\pio_1$ and $\pio_2$, this implies 
that $\tauo$ indeed corresponds to a tensionless domainwall. By the same
token, one should work with $\tilde\calt=\tauo$ in \eqref{simply} in order
to take the holomorphic limit of the infinitesimal invariant at
$\psi=0$.

Via the connection to 4d physics, the existence of a tensionless domainwall 
will lead to singularities in the topological string amplitudes.
This is similar to the appearance of a massless BPS state at the conifold
in the closed string story. The singularity due to a tensionless domainwall
is slightly milder in the sense that at least some of the massless
states appear on the string worldsheet.
In terms of the tensionless domainwall $\tauo$, the leading singularity 
of the $\F gh$ is expected to be
\begin{equation}
\F gh \sim (\tauo)^{2-2g-h}
\end{equation}
This is precisely the singularity we have allowed in our ansatz for the
holomorphic ambiguity.

To quantify the structure of the $\F gh$ around the Gepner point more
explicitly, we use the canonical coordinates and K\"ahler gauge \cite{hkq}
\begin{equation}
s = \frac{\pio_2}{\pio_1} \,,\qquad\qquad
\ee^{-K} = 5^{-3/2} \pio_1
\end{equation}
With these definitions, we obtain the following expansions
\begin{equation}
\eqlabel{gepnerexp}
\begin{split}
\calt_0 &= -\frac{20}{3} s^{3/2} -\frac{4955}{108108} s^{13/2}
-\frac{1007347465}{124547601024} s^{23/2} + \cdots \\[.2cm]
\Delta_{ss} &=   
- 5s^{-1/2} - \frac{4955}{3024}s^{9/2} - 
\frac {1007347465} {1031450112} s^{19/2} + \cdots \\[.2cm]
\F 02_s &=
\frac {103} {50}s^{-1} + \frac {34921} {37800}s^4 + 
\frac {4345923475} {8122669632} s^9 + \cdots \\[.2cm]
\F 11 &= -\frac{67}{150}s^{-3/2} + \frac {4523}{7200} s^{7/2}+ 
\frac {207513043}{1628605440} s^{17/2} + \cdots \\[.2cm]
\F 03 &= -\frac {4616} {9375} s^{-3/2} - \frac {457217} {1181250}s^{7/2} - 
\frac {1069164825109} {5076668520000}s^{17/2} +\cdots
\end{split}
\end{equation}
where we have used the values of the holomorphic ambiguity obtained at
large volume.

\subsection{Conifold expansion}

The expansion around the conifold point, $\psi=1$, is the hardest because it 
cannot be done completely analytically. In the local coordinate
$x= 5^{-5}z^{-1} - 1$, the Picard-Fuchs operator is
\begin{equation}
\callc = (1+x)\call = x(1+x)^4\del_x^4 + 2(1+x)^3(1+3x)\del_x^3+
\frac 15 (1+x)^2(23+35 x)\del_x^2 + (1+x)^2\del_x -\frac{24}{625}
\end{equation}
and has as a basis of solutions \cite{hkq},
\begin{equation}
\eqlabel{notknown}
\begin{split} 
\pic_0 &= 1 + \frac {2} {625} x^3 + \cdots \\
\pic_1 &= x - \frac {3} {10} x^2 + \frac {11} {75}x^3 + \cdots \\
\pic_2 &= x^2 - \frac {23} {30} x^3 + \cdots \\
\pic_3 &= \pic_1 \log(x) +\frac {9} {20} x^2 - \frac {169} {450} x^3 + \cdots
\end{split}
\end{equation}
It is not known analytically what linear combinations of those solutions
correspond to integral (or even real) periods. But the change of basis
between \eqref{notknown} and the large volume symplectic basis 
\eqref{gauge} can be determined numerically \cite{hkq}, and this
knowledge is sufficient for studying the holomorphic limit at the conifold,
both for the closed and for the open string.

From the monodromy around the conifold, it at least follows that
$\pic_1$ is an integral period and that the intersection with the cycle
corresponding to $\pic_3/(2\pi\ii)$ is one. This is enough to conclude that 
the K\"ahler potential has an expansion of the form
\begin{equation}
\ee^{-K(x,\xb)} = \om_0 + \xb \om_1 +\xb\log \xb \pic_1 + O(\xb^2)
\end{equation}
where $\om_0$ and $\om_1$ are some linear combinations of
$\pic_0$, $\pic_1$ and $\pic_2$. Thus, the leading behavior of
the metric at $\xb\to 0$ is a logarithmic divergence
\begin{equation}
G_{x\xb} =\del_x\del_\xb K = -\del_x\Bigl(\frac{\pic_1}{\om_0}\Bigr)\log \xb + O(\xb^0)
\end{equation}
Note that the higher order terms drop out of the Christoffel connection.
Thus, taking the holomorphic limit at the conifold amounts to working 
with
\begin{equation}
\ee^{-K} = \om_0 \,,\qquad t_D = \frac{\pic_1}{\om_0}
\end{equation}

For the open string, the inhomogeneous Picard-Fuchs equation around
the conifold is, \cf, eq.\ \eqref{inhomo}
\begin{equation}
\callc\calt = \frac{15}{16\pi^2} 5^{-5/2} \sqrt{1+x}
\end{equation}
with particular solution $\tauc$ given by
\begin{equation}
4\pi^2 \tauc = \frac{1}{80\sqrt{5}}x^3 - \frac{13}{800\sqrt{5}}x^4
+\frac{5421}{320000\sqrt{5}} x^5
\end{equation}
As for the periods, one can determine numerically precisely which 
solution corresponds to the analytic continuation of $\calt$ from 
\eqref{raw}. But to take the holomorphic limit of the infinitesimal 
invariant, we better find a solution $\tilde\caltc$ which is equivalent 
to $\calt$ modulo real periods such that $C_{xx}^{\;\;\;\xb} 
D_\xb\overline{\tilde\caltc}\to 0$ as $\xb\to 0$ (see the discussion 
around eq.\ \eqref{simply}). Because of the singularity in the metric, 
it is in fact sufficient to have $\tilde\caltc \sim x$ as $x\to 0$. 

At the end, this only leaves one parameter $\alpha$ whose
precise value needs to be determined numerically. We find that the 
requisite solution is
\begin{equation}
\eqlabel{note}
\tilde\caltc = 4\pi^2\tauc + \frac{\alpha}{\sqrt{5}} \pic_2
=\frac{\alpha}{\sqrt{5}} x^2 + \frac{\sqrt{5}(3-184\alpha)}{1200} x^3 
+ \cdots
\end{equation}
with $\alpha\approx -0.002396$. Note that despite appearances, \eqref{note} 
is not a tensionless domainwall, because we have been working modulo real, 
not integral periods.

With all this in hand, we obtain the following expansion of the 
open topological string amplitudes around the conifold:
\begin{equation}
\begin{split}
\F 02_{t_D} &= -\frac{3}{250 t_D} + \frac{21-1280 b_1}{2500} + \cdots \\
\sqrt{5} \F 11 &= -\frac{9}{1000 t_D} + \frac{12633+56500\alpha}{15000} +\cdots\\
\sqrt{5} \F 03 &= -\frac{3}{15625 t_D} + \frac{9423+64000\alpha}{312500}+\cdots
\end{split}
\end{equation}
where $b_1 \approx 0.1641$ is a numerical parameter from \cite{hkq}.

\begin{acknowledgments}
I owe special thanks to Andrew Neitzke for extensive discussions and many
helpful suggestions. I would also like to thank Simeon Hellerman, Manfred 
Herbst, Juan Maldacena, Marcos Mari\~no, David Morrison, Tony Pantev, Jake 
Solomon, J\"org Teschner, and Cumrun Vafa for valuable discussions and 
communications. I thank Albrecht Klemm for bringing ref.\ \cite{schumacher} 
to my attention. This work was supported in part by the Roger Dashen Membership 
at the Institute for Advanced Study and by the NSF under grant number 
PHY-0503584.
\end{acknowledgments}

\end{document}